\documentclass[11pt,a4paper]{article}

\usepackage{verbatim,amsmath,amssymb}
\usepackage{float,color}
\usepackage{geometry}
\usepackage{upgreek}
\usepackage{hyperref}
\usepackage{setspace}
\usepackage{siunitx}
\usepackage{gensymb}
\usepackage{adjustbox}
\usepackage{multirow}
\usepackage[labelfont=bf]{caption,subcaption}
\usepackage{graphicx}
\usepackage[percent]{overpic}
\usepackage{pgfplots}
\usepackage[affil-it]{authblk}
\usepackage{tikz}
\usetikzlibrary{spy,backgrounds}
\pgfplotsset{compat=newest}
\pgfplotsset{scaled y ticks=false}
\graphicspath{{.}{./images/}}
\definecolor{darkblue}{rgb}{0,0,1}
\definecolor{darkgreen}{rgb}{0,0.65,0}
\definecolor{lightblue}{rgb}{0.25,0.5,1}
\definecolor{darkred}{rgb}{0.7,0,0}
\definecolor{orange}{rgb}{1,0.5,0}
\hypersetup{colorlinks=true, breaklinks=true, linkcolor=darkblue, menucolor=darkblue, citecolor=darkblue, urlcolor=darkblue}

\newcommand{\fig}[1]{Fig.~\ref{#1}}								
\newcommand{\figs}[1]{Figs.~\ref{#1}}						
\newcommand{\sect}[1]{Sec.~\ref{#1}}						
					
\newcommand{\tab}[1]{Tab.~\ref{#1}}							
\newcommand{\tabs}[1]{Tabs.~\ref{#1}}					

\newcommand{\bg}{\boldsymbol{g}}
\newcommand{\bx}{\boldsymbol{x}}

\newcommand{\bn}{\boldsymbol{n}}
\newcommand{\btau}{\boldsymbol{\tau}}
\newcommand{\bxi}{\boldsymbol{\xi}}
\newcommand{\ba}{\boldsymbol{a}}

\newcommand{\bN}{\mathbf{N}}
\newcommand{\bC}{\mathbf{C}}
\newcommand{\bB}{\mathbf{B}}
\newcommand{\bI}{\boldsymbol{I}}
\newcommand{\bff}{\mathbf{f}}
\newcommand{\muk}{\mu_\mathrm{b}}
\newcommand{\mus}{\mu_\mathrm{ub}}
\newcommand{\as}{a_\mathrm{s}}
\newcommand{\bs}{b_\mathrm{s}}
\newcommand{\csurf}{\partial_\text{c}\mathcal{B}}
\newcommand{\cbody}{\mathcal{B}}
\newcommand{\cparam}{\mathcal{P}}
\newcommand{\Texp}{T_\mathrm{exp}}
\newcommand{\Tmax}{T_\mathrm{max}}
\newcommand{\Tinf}{T_\infty}
\newcommand{\phio}{\phi_0}
\newcommand{\phiog}{\bar{\phi}_0}
\newcommand{\Eadh}{\bar{w}_\mathrm{adh}}
\newcommand{\errmp}{e^\mathrm{mp}_T}
\newcommand{\errrel}{e^\mathrm{rel}_T}
\newcommand{\dd}{\,\mathrm{d}}
\newcommand{\slrate}{\dot{\bg}_\text{s}}
\newcommand{\ctt}{\boldsymbol{t}_\mathrm{t}}
\newcommand{\ctn}{\boldsymbol{t}_\mathrm{n}}
\newcommand{\ctc}{\boldsymbol{t}_\mathrm{c}}
\newcommand{\ttmax}{t^\mmax_\text{t}}
\newcommand{\thetamax}{\theta_\mmax}
\newcommand{\thetalin}{\theta_\text{lin}}
\newcommand{\gs}{g_\text{s}}
\newcommand{\Eb}{E_\text{b}}
\newcommand{\Gb}{G_\text{b}}
\newcommand{\fs}{f_\text{s}}
\newcommand{\Wdeb}{E_\text{deb}}
\newcommand{\Wfric}{E_\text{fric}}
\newcommand{\Wadh}{W_\text{adh}}
\newcommand{\sigzt}{\sigma_{\theta z}}
\newcommand{\bge}{\bg_\text{e}}
\newcommand{\bgte}{\bg_{\text{et}}}
\newcommand{\bgne}{\bg_\text{en}}

\newcommand{\bgtemax}{\bg_\text{et}^\mmax}

\newcommand{\mmax}{\mathrm{max}}
\newcommand{\etal}{\textit{et al.}}
\newcommand{\lv}{\left\Vert}
\newcommand{\rv}{\right\Vert}

\begin{document}
	\renewcommand*{\thefootnote}{\fnsymbol{footnote}}
%--------------------------------------------------------------------
% Title
%--------------------------------------------------------------------
\title{\Large{\bf{A modified Coulomb's law for the tangential debonding\\ of osseointegrated implants}}}

\author[1,2]{Katharina Immel} 
\author[1]{Thang X. Duong}
\author[3]{Vu-Hieu Nguyen}
\author[2]{\\ Guillaume Ha\"{i}at}
\author[1]{Roger A. Sauer\footnote{Corresponding author, email: sauer@aices.rwth-aachen.de}}

\affil[1]{Aachen Institute for Advanced Study in Computational Engineering Science (AICES), RWTH Aachen University, Templergraben 55, 52056 Aachen, Germany}
\affil[2]{CNRS, Laboratoire Mod\'elisation et Simulation Multi Echelle, MSME UMR 8208 CNRS, 61 Avenue du G\'en\'eral de Gaulle, 94010 Cr\'eteil Cedex, France}
\affil[3]{Universit\'e Paris-Est, Laboratoire Mod\'elisation et Simulation Multi Echelle, MSME UMR 8208 CNRS, 61 Avenue du G\'en\'eral de Gaulle, 94010 Cr\'eteil Cedex, France}
\vspace{-20mm}
\date{}
\maketitle
\vspace{-12mm}
\begin{center}
	{\small Published\footnote[2]{This PDF is the personal version of an article whose final publication is available at \href{https://link.springer.com/article/10.1007\%2Fs10237-019-01272-9}{www.springer.com}} in \textit{Biomechanics and Modeling in Mechanobiology} \href{https://doi.org/10.1007/s10237-019-01272-9}{DOI:~10.1007/s10237-019-01272-9} \\
		Submitted on 17. August 2019, Revised on 16. October 2019, Accepted on 23. November 2019}
\end{center}

%
%--------------------------------------------------------------------
% Abstract and keywords
%--------------------------------------------------------------------
\rule{\linewidth}{.15mm}
{\bf Abstract:}	Cementless implants are widely used in orthopedic and oral surgery. However, debonding-related failure still occurs at the bone-implant interface. 
It remains difficult to predict such implant failure since the underlying osseointegration phenomena are still poorly understood.
Especially in terms of friction and adhesion at the macro-scale, there is a lack of data and reliable models. 
The aim of this work is to present a new friction formulation that can model the tangential contact behavior between osseointegrated implants and bone tissue, with focus on debonding. 
The classical Coulomb's law is combined with a state variable friction law to model a displacement-dependent friction coefficient. 
A smooth state function, based on the sliding distance, is used to model implant debonding. The formulation is implemented in a 3D nonlinear finite element framework, and it is calibrated with experimental data and compared to an analytical model for mode~III cleavage of a coin-shaped, titanium implant~\cite{mathieu2012}. 
Overall, the results show close agreement with the experimental data, especially the peak and the softening part of the torque curve with a relative error of less than \SI{2.25}{\percent}. 
In addition, better estimates of the bone's shear modulus and the adhesion energy are obtained. The proposed model is particularly suitable to account for partial osseointegration, as is also shown.

{\bf Keywords:}	sticking and sliding friction, adhesion, computational contact mechanics, bone-implant interface, state variable friction, finite element method

\vspace{-5mm}
\rule{\linewidth}{.15mm}
%
%--------------------------------------------------------------------
% Introduction
%--------------------------------------------------------------------
\section{Introduction} \label{s:intro}
% motivate topic: implant tability
Endosseous implants are widely used in orthopedic and dental surgery, for instance in total hip and knee arthoplasty or tooth replacement.
Implant stability is a strong determinant of implant success, and it is established through bone growing around and into the rough surface of the implant, a process called \textit{osseointegration}~\cite{albrektsson1981}.
However, debonding-related implant failures still occur and are difficult to anticipate, as bone remodeling and osseointegration phenomena are complex and remain poorly understood.
Although these phenomena include biological, biochemical, and biomechanical factors, the biomechanical properties of the bone-implant interface (BII) are the governing factors of implant stability~\cite{haiat2014,mathieu2011a,mathieu2011b,mathieu2014}.

% stability and influential factors
Two kinds of implant stability can be distinguished: primary (or initial) stability and secondary (or long term) stability (see for example Refs. \cite{kim2008,sul2001}).
Primary stability is achieved during surgery and is mainly governed by interlocking phenomena and bone quality, whereas secondary stability is achieved several weeks or months after surgery, through the formation and maturation of newly formed bone tissue at the BII.
While the evolution of secondary implant stability is governed by complex biochemical processes, the mechanical behavior of the BII remains crucial for the surgical outcome~\cite{gao2019}.
The difficulty of predicting stability arises from the complex nature of the BII, related to i) the implant surface roughness, ii) the non-homogeneous contact between bone and implant, iii) adhesion and friction phenomena, iv) the time-dependent change of periprosthetic bone properties~\cite{mathieu2014} and v) the widely varying load cases during the implant life cycle.
While the implant material, roughness and surface coating are important factors determining implant stability, friction phenomena also play a major role, and in turn depend on the surface roughness and bone quality \cite{shirazi-adl1993}. 

% choice of experiment
Most studies on bone attachment to implants have used push-in or pull-out \textit{in vitro} tests (see for example Refs. \cite{berahmani2015,bishop2014,damm2015,wennerberg2014}).
As the implant geometry influences the test results \cite{branemark1998} and leads to spatially complex, non-uniform, multiaxial stress fields \cite{shirazi1992} and instantaneous crack propagation, using realistic implant geometries makes it difficult to estimate a physically meaningful value for the interfacial mechanical strength.
Therefore, models with a planar BII were designed to minimize the effects of friction and mechanical forces introduced by the geometry \cite{skripitz1999}, see e.g. the experiments on coin-shaped implants by R{\o}nold~\etal~\cite{ronold2002,ronold2003}.
However, the measured pull-out force in these experiments cannot be used to retrieve information about the strength of the interface.
More recently, mode III cleavage experiments applied to coin-shaped implants have been proposed by Mathieu~\etal~\cite{mathieu2012}.
A rotation of the implant with respect to the bone was imposed and the resulting moment was recorded.
This resulted in stable crack propagation, which allows to assess the adhesion energy.
However, the agreement between the analytical and experimental results was only moderate and significant discrepancies were obtained.

% friction in numerical models; motivate presented friction model
The tangential load-displacement behavior at the BII was found to be nonlinear \cite{dammak1997b,rancourt1990,shirazi-adl1993}.
However, numerical analyses of implant stability still only consider either fully bonded, frictionless contact or pure Coulomb's friction (see e.g. Refs.~\cite{chong2010,ghosh2014,pettersen2009}).
While such assumptions may be valid for the assessment of initial stability, prediction of failure of osseointegrated implants requires a more accurate description of the contact behavior and the inclusion of frictional and adhesive effects~\cite{castellani2011,tschegg2011}.

% present friction model
A simple way to model macroscopic friction phenomena that are related to different states of the BII are \textit{state variable friction laws}, introduced by Rice and Ruina~\cite{rice1983,ruina1983}, which were motivated by the experimental findings of Dieterich~\cite{dieterich1978,dieterich1979a}.
These laws focus on the observed phenomena of i) fading memory and steady state, ii) positive dependence on the instantaneous slip rate, and iii) negative dependence on past slip rates.
In general, it is assumed that at any given time, the contact surface has a certain state and the frictional stress only depends on that state, the slip rate and the contact pressure. Similarly, the rate of change for the state only depends on the current state, the slip rate, and the pressure at the analyzed point.
Although these laws have so far been mainly applied in geology and geophysics, one can also interpret the state variable as the osseointegration degree of the BII. 

% aim
The aim of this work is to propose a phenomenological model for the frictional contact behavior of debonding osseointegrated implants.
The classical Coulomb's law is extended from a constant to a varying friction coefficient, that models the transition from an unbroken to a broken state, based on a state variable depending on the total sliding distance of the implant.
While the unbroken state denotes osseointegration and thus the presence of adhesive bonds and a higher friction coefficient, the broken state denotes pure frictional contact behavior of the interface with a lower friction coefficient.
Thus, this model can account for the higher tangential forces observed in osseointegrated implants compared to unbonded implants.

% content
The remainder of this work is structured as follows: \sect{s:model} describes the governing equations and the contact formulation, including the new friction model. In \sect{s:case}, the setup of the mode III cleavage experiment of osseointegrated implant is presented.
Additionally, the corresponding analytical model developed by Mathieu~\etal~\cite{mathieu2012} and the analytical and numerical model based on the proposed friction model of \sect{s:model_modcoulomb} are introduced.
The new analytical and numerical models are calibrated with experimental data and compared to the analytical model from Ref.~\cite{mathieu2012} in \sect{s:results}.
Finally, \sect{s:concl} gives a conclusion and an outlook to possible extensions.
% DONT FORGET TO CHECK IN THE END
%
%--------------------------------------------------------------------
% Modeling
%--------------------------------------------------------------------
\section{Models and Methods} \label{s:model}
This section discusses the governing equations describing the contact behavior, as well as the new friction model.
The resulting equations are discretized within a finite element framework to obtain a numerical solution, which is briefly summarized in Appendix \ref{s:appndx_contact}.
The readers are referred to Duong and Sauer~\cite{duong2019} and Sauer and De Lorenzis~\cite{sauer2013,sauer2015} for a more detailed derivation of the considered contact formulation.
	%
	%--------------------------------------------------------------------
	% Material law
	%-------------------------------------------------------------------	
	\subsection{Material laws} \label{s:model_material}
	Throughout this work, a hyper-elastic Neo-Hookean material model for both, bone and implant, is used. According to Zienkiewicz and Taylor~\cite{zienkiewicz2005} the stress-strain relation for the Cauchy stress $\boldsymbol{\sigma}$ of this model is given by
	\begin{equation} \label{eq:neohooke}
		\boldsymbol{\sigma} = \frac{\Lambda}{J} \left( \mathrm{ln} J \right) \bI + \frac{G}{J} \left( \boldsymbol{b} - \bI \right), 
	\end{equation}
	with the volume change $J$, the identity tensor $\bI$, and the left Cauchy-Green tensor $\boldsymbol{b}$. The Lamé parameters $G$ (shear modulus) and $\Lambda$ can be expressed in terms of Young's modulus $E$ and Poisson ration $\nu$, by
	\begin{equation}\label{eq:young}
		G = \frac{E}{2(1+\nu)} \quad \mathrm{and} \quad \Lambda = \frac{2G \nu}{1 - 2\nu}.
	\end{equation}
	%
	%--------------------------------------------------------------------
	% Friction laws
	%-------------------------------------------------------------------	
	\subsection{Friction laws} \label{s:model_fric}
	Given two 3D bodies $\cbody_1$ and $\cbody_2$ and their contact surface $\csurf$, the contact traction $\ctc$ can be decomposed into a normal and tangential component, i.e.
	\begin{equation}
		\ctc = \ctn - \ctt.
	\end{equation}
	The magnitude of the normal traction is equal to the normal pressure, i.e.
	\begin{equation} \label{eq:normal_traction}
		\lv \ctn \rv = p. 
	\end{equation} 
	For frictional contact, the tangential traction is determined by the behavior during sticking and sliding, and the distinction between these two states is based on a slip criterion of the form
	\begin{equation}
		\fs \begin{cases}
				<0, & \text{for sticking},\\
				=0, & \text{for sliding}.
		\end{cases}
	\end{equation}
	It can be formulated as
	\begin{equation}
		\fs = \lv \ctt \rv - \ttmax , \quad	\lv \ctt \rv \le \ttmax,
	\end{equation}
	where $\ttmax>0$ is the maximal tangential contact traction. For sticking, the tangential traction is defined by the constraint that no relative tangential motion occurs, while for sliding, the tangential traction is defined by a sliding law.
		%
		%--------------------------------------------------------------------
		% Pure Coulomb
		%-------------------------------------------------------------------
		\subsubsection{Coulomb's law} \label{s:model_coulomb}
		For classical Coulomb's friction, the slip function $\fs$ can be defined as
		\begin{equation}
			\fs = \lv \ctt \rv - \mu p,
		\end{equation}
		where $\mu$ denotes the (constant) friction coefficient. Then, the tangential sliding traction is given by
		\begin{equation} \label{eq:coulomb}
			\ctt = - \mu p \frac{\slrate}{\lv \slrate \rv}, 
		\end{equation}
		with $\slrate$ denoting the relative sliding velocity between the two bodies in contact, see Appendix~\ref{s:appndx_contact}.
		%
		%--------------------------------------------------------------------
		% Rate-state friction
		%-------------------------------------------------------------------
		\subsubsection{State variable friction laws} \label{s:model_svfl}
		In so called \textit{rate--state} or \textit{state variable} friction laws (see, e.g. Rice and Ruina \cite{rice1983,ruina1983}) it is assumed, that at any given time, any point $\bx$ on the contact surface has a state $\phi=\phi(\bx,t)$.
		The tangential contact traction then depends in general on the contact pressure $p$, the sliding rate $\dot{\gs}=\lv\slrate\rv$, and the state variable $\phi$, i.e.
		\begin{equation}
			\ctt = \ctt(p,\dot{\gs },\phi).
		\end{equation}
		For a given $\bx$, the state $\phi$ is assumed to be a continuous function of $\dot{\gs}(t)$. Moreover, the rate of change of $\phi$ at $\bx$ generally only depends on the pressure, the sliding rate and the instantaneous state of this point, i.e.
		\begin{equation} \label{eq:svfl_evolution}
			\dot{\phi} = \text{F}(p,\dot{\gs},\phi). 
		\end{equation}
		Since $\phi$ does not depend on the state at other points, Eq.~\eqref{eq:svfl_evolution} is a local evolution law.
		State variable friction laws are able to model a change in the frictional contact traction due to past states (referred to as \textit{memory} \cite{rice1983,ruina1983}).
		They are also able to model an asymptotic approach to steady state sliding when $\dot{\gs}$ becomes constant~\cite{rice1983}.
		%
		%--------------------------------------------------------------------
		% Modified Coulomb's law
		%-------------------------------------------------------------------
		\subsubsection{Modified Coulomb's friction law} \label{s:model_modcoulomb}
		In contrast to a constant friction coefficient as is used in the classical Coulomb's law \eqref{eq:coulomb}, we propose to model $\mu$ as a function of the scalar state variable $\phi$, as
		\begin{equation}
			\mu := \mu(\phi) = \phi \, \mus + \left( 1 - \phi \right) \muk, \label{eq:mod_mu}
		\end{equation}
		where $\mus$ and $\muk$ are the friction coefficient for the unbroken (initial) and broken state, respectively, that are weighted according to a state variable. As this is a local friction model, where $\phi$ and thus $\mu$ can change pointwise, it allows for the description of locally varying bonding states, such as occur in crack propagation and partial osseointegration.
		
\begin{figure}[H]
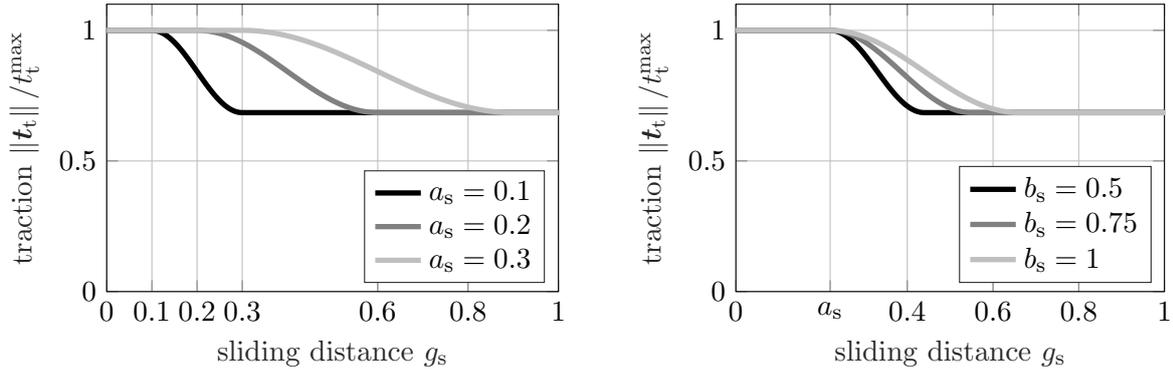

			\centering
			\begin{subfigure}[t]{0.48\textwidth}
				\input{./images/law_tt_gs_4.tex}
				\caption{Variation of the sliding threshold $\as$ for $\bs=1$.}
				\label{img:tt_gs_as}
			\end{subfigure}
			\hfill
			\begin{subfigure}[t]{0.48\textwidth}
				\input{./images/law_tt_gs_1.tex}
				\caption{Variation of the transition factor $\bs$ for $\as=0.22$.}
				\label{img:tt_gs_bs}
			\end{subfigure}
			\caption{Modified Coulomb's law: Tangential contact traction $\lv \ctt \rv / \ttmax$ as a function of the sliding distance $\gs$ for varying sliding threshold $\as$ and transition factor $\bs$ for $\phio=1$.}
			\label{img:modcoulomb}
		\end{figure}

		According to Eq.~\eqref{eq:mod_mu}, the state variable $\phi$ determines whether a point is in an unbroken, partially broken or fully broken state.
		Here, $\phi$ is considered to depend on the accumulated sliding distance
		\begin{equation}
			\gs = \displaystyle \int_0^t \dot{\gs} \, \dd t  \label{eq:gs}
		\end{equation}
		at a certain point $\bx$, according to the smooth function\footnote{In principle Eq.~\eqref{eq:phi} can be brought into the form of Eq.~\eqref{eq:svfl_evolution} if a dot is applied to Eq.~\eqref{eq:phi} and then $\gs$ is eliminated by the inverse function of \eqref{eq:phi}.}
		\begin{equation} \label{eq:phi}
		\displaystyle
			\phi(\gs)= \phio \cdot \begin{cases}
							1 & \text{if} \quad \frac{\gs}{\as} < 1, \\
							\frac{1}{2} - \frac{1}{2} \mathrm{sin} \left( \frac{\pi}{2\bs} \left(\frac{\gs}{\as} - \bs - 1 \right) \right) 
							& \text{if}  \quad 1 \le \frac{\gs}{\as} \le 1+2\bs,\\
							0 & \text{if} \quad \frac{\gs}{\as} > 1+2\bs.
						\end{cases}
		\end{equation}
		This function was designed such that it captures the experimental behavior shown in \sect{s:res_torque} (see \fig{img:torque}).
		Equation \eqref{eq:phi} depends on the initial bonding state $\phio=[0,1]$, where $\phio(\bx)=0$ denotes no initial bonding and $\phio(\bx)=1$ represents full initial bonding.\footnote{In principle, $\phio$ follows from an evolution law that describes the healing/osseointegration process ($\phio$ increasing from 0 to 1). However, this is not considered here. Here, only the (further) evolution of $\phi$ during debonding ($\phi$ decreasing from 1 to 0) is studied.}
		The bonding state variable indicates that for $\phi(\bx)>0$ tangential contact at $\bx$ is governed by the proposed friction law \eqref{eq:mod_mu}, while for $\phi(\bx)=0$ it is governed by classical Coulomb's law \eqref{eq:coulomb} with $\mu=\muk$.
		This definition results in three possible states for every point on the contact surface: fully bonded ($\phi=1$), debonding ($0<\phi<1$) and fully debonded/sliding ($\phi=0$), which is illustrated in Figures \ref{img:tt_gs_as} and \ref{img:tt_gs_bs}.
		The parameter $\as$ represents the sliding threshold up to which tangential adhesion takes effect ($\mu=\mus$ for $\gs\le\as$), while $\bs$ defines the size of the transition zone between the two friction coefficients.
		This implies, that up to a sliding length of $\gs=\as$ we have a higher resistance to tangential displacement, similar to the effect of adhesion.
		After the sliding distance $\as$ is reached, the friction coefficient starts to decrease to $\mu=\muk$, corresponding to the sliding of a fully debonded body. The sliding distance needed for a certain point on the contact surface to fully debond is then controlled by the parameter $\bs$. 
		%--------------------------------------------------------------------
		%--------------------------------------------------------------------
		For simple cases, this model can be solved analytically, as presented in \sect{s:case_ana_new}. However, for the complex geometries of endosseous implants, the nonlinear and anisotropic behavior of bone tissue and loading conditions inside the human body may require a numerical solution. This becomes particularly important when considering inhomogeneous initial bonding (where $\phio(\bx)\neq\text{const} \, \forall \bx)$, due to imperfect or partial osseointegration, as presented in \sect{s:res_osseo}.
%	
%--------------------------------------------------------------------
% Test case
%--------------------------------------------------------------------
\section{Mode III debonding of osseointegrated coin-shaped implants} \label{s:case}
To calibrate the friction model described above, it is applied to the mode III debonding of an osseointegrated implant. The test case is originally presented by Mathieu~\etal~\cite{mathieu2012} (in the following also referred to as the reference study). First, a short summary of the experimental and analytical results from the reference study are given. Second, a new analytical model based on the proposed friction model of \sect{s:model_modcoulomb} is introduced. Third, the numerical setup and parameters are presented. 
	%
	%--------------------------------------------------------------------
	% Experimental setup
	%--------------------------------------------------------------------	
	\subsection{Experimental setup} \label{s:case_experiment}
	In Mathieu~\etal~\cite{mathieu2012}, two coin-shaped implants made of titanium alloy (Ti-6Al-4V), with a radius $R=2.5$ \si{\milli\m} and a height $H=3$ \si{\milli\m}, were implanted into the tibiae of a rabbit and left \textit{in vivo} during seven weeks. Polytetrafluoroethylene (PTFE) caps were placed around the implants, to ensure that bone in-growth only occurred at the bottom of the cylindrical implants. After seven weeks, the rabbit was sacrificed and the bone samples including the implants were extracted and conserved. Then, mode III cleavage experiments were carried out. The bone specimen was rigidly fixed to minimize the remaining normal force. The implant was fixed by a chuck screwed to a torque meter. Then, a \ang{10} rotation with a constant rotation speed of \ang{0.01}~\si{\per\second} was imposed. Finally, the torque and rotation angle as a function of time were extracted via post processing. 
	%--------------------------------------------------------------------
	\begin{figure}[H]
		\centering
		\begin{subfigure}[b]{0.48\textwidth}
			\centering
			\def\svgwidth{0.95\textwidth}
			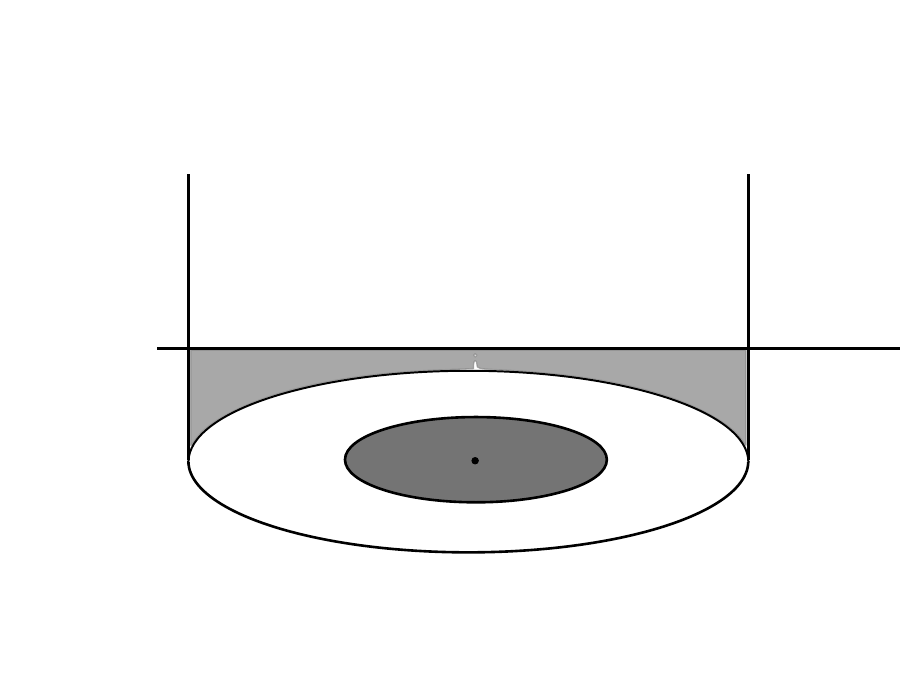
		%	\vspace{-4.5cm}
			\caption{Schematic representation of the setup and the crack propagation in the uniformly bonded case.}
			\label{img:exp}
		\end{subfigure}
		\hfill
		\begin{subfigure}[b]{0.48\textwidth}
			\centering
			\includegraphics[width=\textwidth]{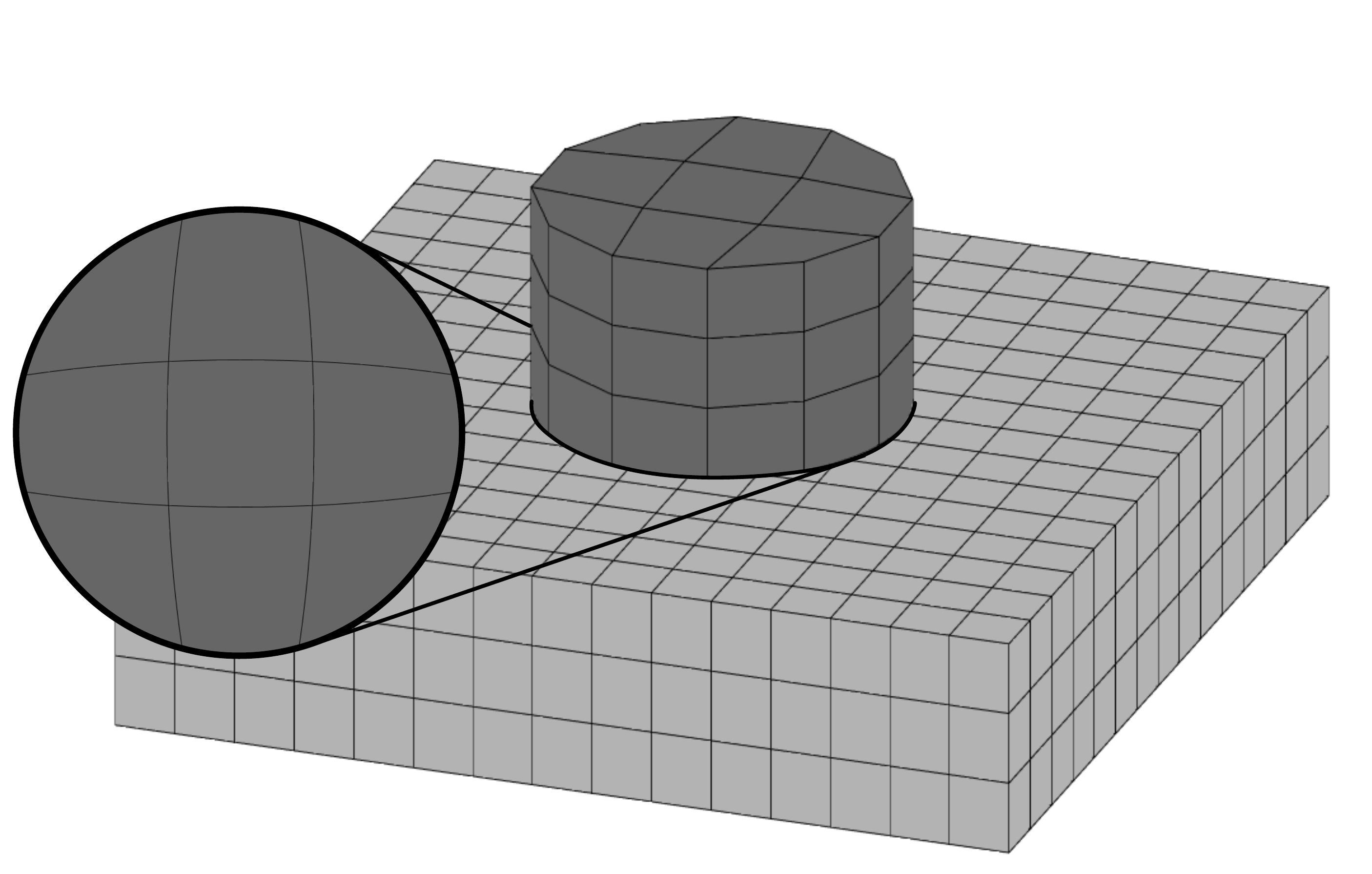}
		%	\vspace{0.6mm}
			\caption{NURBS-enriched finite element mesh showing an enlargement of the contact surface.}
			\label{img:exp2}
		\end{subfigure}
		\caption{Illustrations for the analytical and numerical setup.}
		\label{img:setup}
	\end{figure}
	%--------------------------------------------------------------------
	%
	%--------------------------------------------------------------------
	% Analytical model
	%--------------------------------------------------------------------
	\subsection{Analytical model of Mathieu et al. \cite{mathieu2012}} \label{s:case_ana}
	To explain the experimental results, Mathieu~\etal~\cite{mathieu2012} developed an analytical model that couples crack propagation with friction, based on Chateauminois~\etal~\cite{chateauminois2010}. The model assumes that the crack starts from the outside and propagates purely in a circular manner and in mode III to the center of the contact surface, using a Dugdale crack model, without any normal force. In this description, the interfacial forces are supposed to be constant up to a given separation distance between the surfaces. Let $R$ be the radius of the implant and the initial contact area, then $c$ defines the radius of the current crack (see \fig{img:exp}), corresponding to the twisting angle
	\begin{equation}
		\theta = \sqrt{\frac{\pi \Eadh}{4c G}} + \frac{\tau_0}{2G} \mathrm{cosh}^{-1} \left( \frac{R}{c} \right),
	\end{equation} 
	where $\Eadh$ denotes the adhesion energy. $c$ separates the contact area into an adhering/sticking region for $r<c$ and a debonded/sliding region for $c<r<R$. In the sliding region, the orthoradial shear stress $\sigzt$ is assumed to be constant, according to
	\begin{equation}
		\sigzt(r)=\tau_0 \quad \mathrm{for} \,\,c<r<R, \label{eq:ana_old1}
	\end{equation}
	with constant $\tau_0 = 3\Tinf/2\pi R^3$, where $\Tinf$ corresponds to the torque for a rotation angle equal to infinity (i.e., where the surface is fully debonded), and $z$ denotes the axial direction. In the sticking region, the orthoradial stress becomes
	\begin{equation} \label{eq:ana_old2}
		\sigzt(r) = \frac{2}{\pi} \left[ \sqrt{ \frac{\pi G \Eadh}{c}} \frac{r}{\sqrt{c^2-r^2}} + \tau_0\,\mathrm{sin}^{-1} \left( \sqrt{ \frac{ \left( \frac{R}{c} \right)^2-1}{\left( \frac{R}{r} \right)^2-1}} \right) \right] \mathrm{for} \,\,r < c.
	\end{equation}
	%
	%--------------------------------------------------------------------
	% New analytical solution
	%--------------------------------------------------------------------
	\subsection{New analytical model with modified Coulomb's friction} \label{s:case_ana_new}
	Given the modified Coulomb's law in \sect{s:model_modcoulomb}, a new analytical model can be derived. Here, only the key aspects of the new analytical model are presented. The complete derivation can be found in Appendix \ref{s:appndx_ana_new}.
	
	Due to symmetry, the tangential traction component $\sigzt$ is distributed radially symmetric along the BII, while the radial traction component $\sigma_{rz}$ is zero (as in the model presented in \sect{s:case_ana}). We assume both bodies to be linear elastic and the normal contact pressure $p$ to be distributed homogeneously along the contact surface. We define the hyperbola
	\begin{equation} \label{eq:rcrit}
		c(\theta) := \frac{\thetalin}{\theta}R
	\end{equation}
	to be the function of the critical radius for the stick/slip transition, such that $r < c$ denotes the sticking region and $c \ge r \ge R$ denotes the sliding region. Furthermore, we define $\thetalin$ as the limit for which the tangential traction will be a linear function of the implant radius and $\thetamax$ is the rotation angle for which the whole contact surface starts sliding (see \fig{img:rcrit}). It also marks the location of the maximal torque $\Tmax$ in the torque-angle-curve (see e.g. \fig{img:torque_a}). 

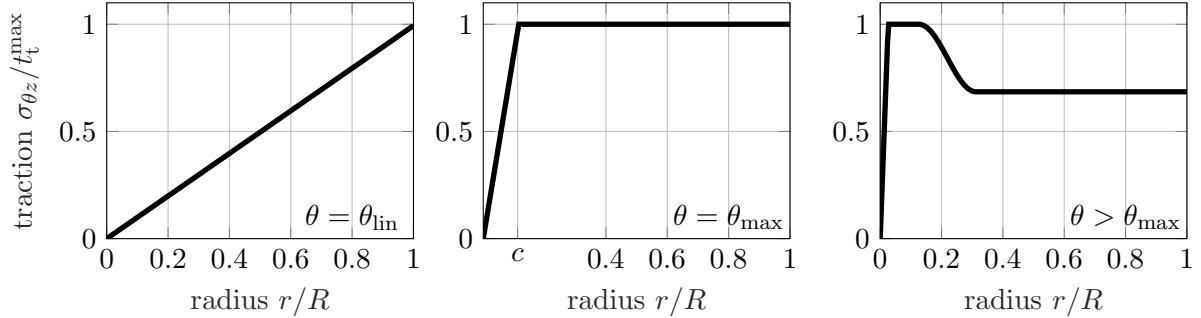
\begin{figure}[H]
		\centering
		\begin{subfigure}[t]{0.32\textwidth}
\definecolor{mycolor1}{rgb}{0.00000,0.44700,0.74100}%
\definecolor{mycolor2}{rgb}{0.85000,0.32500,0.09800}%
\begin{tikzpicture}

\begin{axis}[%
width=1.589in,
height=1.231in,
scale only axis,
xmin=0,
xmax=1,
xlabel style={font=\color{white!15!black}},
xlabel={radius $r/R$},
ymin=0,
ymax=1.1,
ylabel style={font=\color{white!15!black}},
ylabel={traction \linebreak $\sigzt/\ttmax$},
axis background/.style={fill=white},
xmajorgrids,
ymajorgrids,
legend pos=south east,
legend style={legend cell align=left, align=left, draw=white!15!black}
]
\addplot [color=black, line width=2.0pt]
  table[row sep=crcr]{%
0	0\\
0.004	0.00397377830548895\\
0.008	0.00794755661097791\\
0.012	0.0119213349164669\\
0.016	0.0158951132219558\\
0.02	0.0198688915274448\\
0.024	0.0238426698329337\\
0.028	0.0278164481384227\\
0.032	0.0317902264439116\\
0.036	0.0357640047494006\\
0.04	0.0397377830548895\\
0.044	0.0437115613603785\\
0.048	0.0476853396658674\\
0.052	0.0516591179713564\\
0.056	0.0556328962768453\\
0.06	0.0596066745823343\\
0.064	0.0635804528878233\\
0.068	0.0675542311933122\\
0.072	0.0715280094988011\\
0.076	0.0755017878042901\\
0.08	0.0794755661097791\\
0.084	0.083449344415268\\
0.088	0.087423122720757\\
0.092	0.0913969010262459\\
0.096	0.0953706793317349\\
0.1	0.0993444576372238\\
0.104	0.103318235942713\\
0.108	0.107292014248202\\
0.112	0.111265792553691\\
0.116	0.11523957085918\\
0.12	0.119213349164669\\
0.124	0.123187127470158\\
0.128	0.127160905775647\\
0.132	0.131134684081135\\
0.136	0.135108462386624\\
0.14	0.139082240692113\\
0.144	0.143056018997602\\
0.148	0.147029797303091\\
0.152	0.15100357560858\\
0.156	0.154977353914069\\
0.16	0.158951132219558\\
0.164	0.162924910525047\\
0.168	0.166898688830536\\
0.172	0.170872467136025\\
0.176	0.174846245441514\\
0.18	0.178820023747003\\
0.184	0.182793802052492\\
0.188	0.186767580357981\\
0.192	0.19074135866347\\
0.196	0.194715136968959\\
0.2	0.198688915274448\\
0.204	0.202662693579937\\
0.208	0.206636471885426\\
0.212	0.210610250190915\\
0.216	0.214584028496403\\
0.22	0.218557806801892\\
0.224	0.222531585107381\\
0.228	0.22650536341287\\
0.232	0.230479141718359\\
0.236	0.234452920023848\\
0.24	0.238426698329337\\
0.244	0.242400476634826\\
0.248	0.246374254940315\\
0.252	0.250348033245804\\
0.256	0.254321811551293\\
0.26	0.258295589856782\\
0.264	0.262269368162271\\
0.268	0.26624314646776\\
0.272	0.270216924773249\\
0.276	0.274190703078738\\
0.28	0.278164481384227\\
0.284	0.282138259689716\\
0.288	0.286112037995205\\
0.292	0.290085816300694\\
0.296	0.294059594606183\\
0.3	0.298033372911671\\
0.304	0.30200715121716\\
0.308	0.305980929522649\\
0.312	0.309954707828138\\
0.316	0.313928486133627\\
0.32	0.317902264439116\\
0.324	0.321876042744605\\
0.328	0.325849821050094\\
0.332	0.329823599355583\\
0.336	0.333797377661072\\
0.34	0.337771155966561\\
0.344	0.34174493427205\\
0.348	0.345718712577539\\
0.352	0.349692490883028\\
0.356	0.353666269188517\\
0.36	0.357640047494006\\
0.364	0.361613825799495\\
0.368	0.365587604104984\\
0.372	0.369561382410473\\
0.376	0.373535160715962\\
0.38	0.377508939021451\\
0.384	0.38148271732694\\
0.388	0.385456495632428\\
0.392	0.389430273937917\\
0.396	0.393404052243406\\
0.4	0.397377830548895\\
0.404	0.401351608854384\\
0.408	0.405325387159873\\
0.412	0.409299165465362\\
0.416	0.413272943770851\\
0.42	0.41724672207634\\
0.424	0.421220500381829\\
0.428	0.425194278687318\\
0.432	0.429168056992807\\
0.436	0.433141835298296\\
0.44	0.437115613603785\\
0.444	0.441089391909274\\
0.448	0.445063170214763\\
0.452	0.449036948520252\\
0.456	0.453010726825741\\
0.46	0.45698450513123\\
0.464	0.460958283436719\\
0.468	0.464932061742207\\
0.472	0.468905840047696\\
0.476	0.472879618353185\\
0.48	0.476853396658674\\
0.484	0.480827174964163\\
0.488	0.484800953269652\\
0.492	0.488774731575141\\
0.496	0.49274850988063\\
0.5	0.496722288186119\\
0.504	0.500696066491608\\
0.508	0.504669844797097\\
0.512	0.508643623102586\\
0.516	0.512617401408075\\
0.52	0.516591179713564\\
0.524	0.520564958019053\\
0.528	0.524538736324542\\
0.532	0.528512514630031\\
0.536	0.53248629293552\\
0.54	0.536460071241009\\
0.544	0.540433849546498\\
0.548	0.544407627851986\\
0.552	0.548381406157475\\
0.556	0.552355184462964\\
0.56	0.556328962768453\\
0.564	0.560302741073942\\
0.568	0.564276519379431\\
0.572	0.56825029768492\\
0.576	0.572224075990409\\
0.58	0.576197854295898\\
0.584	0.580171632601387\\
0.588	0.584145410906876\\
0.592	0.588119189212365\\
0.596	0.592092967517854\\
0.6	0.596066745823343\\
0.604	0.600040524128832\\
0.608	0.604014302434321\\
0.612	0.60798808073981\\
0.616	0.611961859045299\\
0.62	0.615935637350788\\
0.624	0.619909415656277\\
0.628	0.623883193961766\\
0.632	0.627856972267255\\
0.636	0.631830750572743\\
0.64	0.635804528878232\\
0.644	0.639778307183721\\
0.648	0.64375208548921\\
0.652	0.647725863794699\\
0.656	0.651699642100188\\
0.66	0.655673420405677\\
0.664	0.659647198711166\\
0.668	0.663620977016655\\
0.672	0.667594755322144\\
0.676	0.671568533627633\\
0.68	0.675542311933122\\
0.684	0.679516090238611\\
0.688	0.6834898685441\\
0.692	0.687463646849589\\
0.696	0.691437425155078\\
0.7	0.695411203460567\\
0.704	0.699384981766056\\
0.708	0.703358760071545\\
0.712	0.707332538377034\\
0.716	0.711306316682523\\
0.72	0.715280094988011\\
0.724	0.719253873293501\\
0.728	0.723227651598989\\
0.732	0.727201429904478\\
0.736	0.731175208209967\\
0.74	0.735148986515456\\
0.744	0.739122764820945\\
0.748	0.743096543126434\\
0.752	0.747070321431923\\
0.756	0.751044099737412\\
0.76	0.755017878042901\\
0.764	0.75899165634839\\
0.768	0.762965434653879\\
0.772	0.766939212959368\\
0.776	0.770912991264857\\
0.78	0.774886769570346\\
0.784	0.778860547875835\\
0.788	0.782834326181324\\
0.792	0.786808104486813\\
0.796	0.790781882792302\\
0.8	0.794755661097791\\
0.804	0.79872943940328\\
0.808	0.802703217708769\\
0.812	0.806676996014257\\
0.816	0.810650774319746\\
0.82	0.814624552625235\\
0.824	0.818598330930724\\
0.828	0.822572109236213\\
0.832	0.826545887541702\\
0.836	0.830519665847191\\
0.84	0.83449344415268\\
0.844	0.838467222458169\\
0.848	0.842441000763658\\
0.852	0.846414779069147\\
0.856	0.850388557374636\\
0.86	0.854362335680125\\
0.864	0.858336113985614\\
0.868	0.862309892291103\\
0.872	0.866283670596592\\
0.876	0.870257448902081\\
0.88	0.87423122720757\\
0.884	0.878205005513059\\
0.888	0.882178783818548\\
0.892	0.886152562124037\\
0.896	0.890126340429525\\
0.9	0.894100118735014\\
0.904	0.898073897040503\\
0.908	0.902047675345992\\
0.912	0.906021453651481\\
0.916	0.90999523195697\\
0.92	0.913969010262459\\
0.924	0.917942788567948\\
0.928	0.921916566873437\\
0.932	0.925890345178926\\
0.936	0.929864123484415\\
0.94	0.933837901789904\\
0.944	0.937811680095393\\
0.948	0.941785458400882\\
0.952	0.945759236706371\\
0.956	0.94973301501186\\
0.96	0.953706793317349\\
0.964	0.957680571622838\\
0.968	0.961654349928327\\
0.972	0.965628128233815\\
0.976	0.969601906539304\\
0.98	0.973575684844793\\
0.984	0.977549463150282\\
0.988	0.981523241455771\\
0.992	0.98549701976126\\
0.996	0.989470798066749\\
1	0.993444576372238\\
};
\draw[color=black] (0.8, 0.1) node {$\theta=\theta_\text{lin}$}; 
\end{axis}

\end{tikzpicture}%
			\label{img:thetalin}
		\end{subfigure}
		\hfill
		\begin{subfigure}[t]{0.32\textwidth}
\definecolor{mycolor1}{rgb}{0.00000,0.44700,0.74100}%
\definecolor{mycolor2}{rgb}{0.85000,0.32500,0.09800}%
\begin{tikzpicture}

\begin{axis}[%
width=1.589in,
height=1.231in,
scale only axis,
xmin=0,
xmax=1,
xtick={0.112,0.4, 0.6,0.8,1},
xticklabels={$\large c$,0.4,0.6,0.8,1},
xlabel style={font=\color{white!15!black}},
xlabel={radius $r/R$},
ymin=0,
ymax=1.1,
ylabel style={font=\color{white!15!black}},
axis background/.style={fill=white},
xmajorgrids,
ymajorgrids,
legend pos=south east,
legend style={legend cell align=left, align=left, draw=white!15!black}
]
\addplot [color=black, line width=2.0pt]
  table[row sep=crcr]{%
0	0\\
0.004	0.0345413037323271\\
0.008	0.0690826074646541\\
0.012	0.103623911196981\\
0.016	0.138165214929308\\
0.02	0.172706518661635\\
0.024	0.207247822393962\\
0.028	0.241789126126289\\
0.032	0.276330429858616\\
0.036	0.310871733590943\\
0.04	0.34541303732327\\
0.044	0.379954341055598\\
0.048	0.414495644787925\\
0.052	0.449036948520252\\
0.056	0.483578252252579\\
0.06	0.518119555984906\\
0.064	0.552660859717233\\
0.068	0.58720216344956\\
0.072	0.621743467181887\\
0.076	0.656284770914214\\
0.08	0.690826074646541\\
0.084	0.725367378378868\\
0.088	0.759908682111195\\
0.092	0.794449985843522\\
0.096	0.828991289575849\\
0.1	0.863532593308176\\
0.104	0.898073897040504\\
0.108	0.93261520077283\\
0.112	0.967156504505158\\
0.116	1\\
0.12	1\\
0.124	1\\
0.128	1\\
0.132	1\\
0.136	1\\
0.14	1\\
0.144	1\\
0.148	1\\
0.152	1\\
0.156	1\\
0.16	1\\
0.164	1\\
0.168	1\\
0.172	1\\
0.176	1\\
0.18	1\\
0.184	1\\
0.188	1\\
0.192	1\\
0.196	1\\
0.2	1\\
0.204	1\\
0.208	1\\
0.212	1\\
0.216	1\\
0.22	1\\
0.224	1\\
0.228	1\\
0.232	1\\
0.236	1\\
0.24	1\\
0.244	1\\
0.248	1\\
0.252	1\\
0.256	1\\
0.26	1\\
0.264	1\\
0.268	1\\
0.272	1\\
0.276	1\\
0.28	1\\
0.284	1\\
0.288	1\\
0.292	1\\
0.296	1\\
0.3	1\\
0.304	1\\
0.308	1\\
0.312	1\\
0.316	1\\
0.32	1\\
0.324	1\\
0.328	1\\
0.332	1\\
0.336	1\\
0.34	1\\
0.344	1\\
0.348	1\\
0.352	1\\
0.356	1\\
0.36	1\\
0.364	1\\
0.368	1\\
0.372	1\\
0.376	1\\
0.38	1\\
0.384	1\\
0.388	1\\
0.392	1\\
0.396	1\\
0.4	1\\
0.404	1\\
0.408	1\\
0.412	1\\
0.416	1\\
0.42	1\\
0.424	1\\
0.428	1\\
0.432	1\\
0.436	1\\
0.44	1\\
0.444	1\\
0.448	1\\
0.452	1\\
0.456	1\\
0.46	1\\
0.464	1\\
0.468	1\\
0.472	1\\
0.476	1\\
0.48	1\\
0.484	1\\
0.488	1\\
0.492	1\\
0.496	1\\
0.5	1\\
0.504	1\\
0.508	1\\
0.512	1\\
0.516	1\\
0.52	1\\
0.524	1\\
0.528	1\\
0.532	1\\
0.536	1\\
0.54	1\\
0.544	1\\
0.548	1\\
0.552	1\\
0.556	1\\
0.56	1\\
0.564	1\\
0.568	1\\
0.572	1\\
0.576	1\\
0.58	1\\
0.584	1\\
0.588	1\\
0.592	1\\
0.596	1\\
0.6	1\\
0.604	1\\
0.608	1\\
0.612	1\\
0.616	1\\
0.62	1\\
0.624	1\\
0.628	1\\
0.632	1\\
0.636	1\\
0.64	1\\
0.644	1\\
0.648	1\\
0.652	1\\
0.656	1\\
0.66	1\\
0.664	1\\
0.668	1\\
0.672	1\\
0.676	1\\
0.68	1\\
0.684	1\\
0.688	1\\
0.692	1\\
0.696	1\\
0.7	1\\
0.704	1\\
0.708	1\\
0.712	1\\
0.716	1\\
0.72	1\\
0.724	1\\
0.728	1\\
0.732	1\\
0.736	1\\
0.74	1\\
0.744	1\\
0.748	1\\
0.752	1\\
0.756	1\\
0.76	1\\
0.764	1\\
0.768	1\\
0.772	1\\
0.776	1\\
0.78	1\\
0.784	1\\
0.788	1\\
0.792	1\\
0.796	1\\
0.8	1\\
0.804	1\\
0.808	1\\
0.812	1\\
0.816	1\\
0.82	1\\
0.824	1\\
0.828	1\\
0.832	1\\
0.836	1\\
0.84	1\\
0.844	1\\
0.848	1\\
0.852	1\\
0.856	1\\
0.86	1\\
0.864	1\\
0.868	1\\
0.872	1\\
0.876	1\\
0.88	1\\
0.884	1\\
0.888	1\\
0.892	1\\
0.896	1\\
0.9	1\\
0.904	1\\
0.908	1\\
0.912	1\\
0.916	1\\
0.92	1\\
0.924	1\\
0.928	1\\
0.932	1\\
0.936	1\\
0.94	1\\
0.944	1\\
0.948	1\\
0.952	1\\
0.956	1\\
0.96	1\\
0.964	1\\
0.968	1\\
0.972	1\\
0.976	1\\
0.98	1\\
0.984	1\\
0.988	1\\
0.992	1\\
0.996	1\\
1	1\\
};
\draw[color=black] (0.8, 0.1) node {$\theta=\thetamax$}; 
\end{axis}
\end{tikzpicture}%
			\label{img:thetamax}
		\end{subfigure}
		\begin{subfigure}[t]{0.32\textwidth}
\definecolor{mycolor1}{rgb}{0.00000,0.44700,0.74100}%
\definecolor{mycolor2}{rgb}{0.85000,0.32500,0.09800}%
\begin{tikzpicture}

\begin{axis}[%
width=1.589in,
height=1.231in,
scale only axis,
xmin=0,
xmax=1,
xlabel style={font=\color{white!15!black}},
xlabel={radius $r/R$},
ymin=0,
ymax=1.1,
ylabel style={font=\color{white!15!black}},
axis background/.style={fill=white},
xmajorgrids,
ymajorgrids,
legend pos=south east,
legend style={legend cell align=left, align=left, draw=white!15!black}
]
\addplot [color=black, line width=2.0pt]
  table[row sep=crcr]{%
0	0\\
0.004	0.15283762713419\\
0.008	0.305675254268381\\
0.012	0.458512881402572\\
0.016	0.611350508536762\\
0.02	0.764188135670952\\
0.024	0.917025762805143\\
0.028	1\\
0.032	1\\
0.036	1\\
0.04	1\\
0.044	1\\
0.048	1\\
0.052	1\\
0.056	1\\
0.06	1\\
0.064	1\\
0.068	1\\
0.072	1\\
0.076	1\\
0.08	1\\
0.084	1\\
0.088	1\\
0.092	1\\
0.096	1\\
0.1	1\\
0.104	1\\
0.108	1\\
0.112	1\\
0.116	1\\
0.12	1\\
0.124	1\\
0.128	0.999912365588528\\
0.132	0.999201406067288\\
0.136	0.997779195295779\\
0.14	0.995652187129087\\
0.144	0.992830033724796\\
0.148	0.989325541742419\\
0.152	0.985154614228022\\
0.156	0.980336178447733\\
0.16	0.974892099997639\\
0.164	0.968847083579823\\
0.168	0.96222856089481\\
0.172	0.955066566159164\\
0.176	0.947393599813112\\
0.18	0.939244481036686\\
0.184	0.93065618974365\\
0.188	0.921667698770222\\
0.192	0.912319797020106\\
0.196	0.902654904368392\\
0.2	0.892716879164259\\
0.204	0.882550819206021\\
0.208	0.872202857091678\\
0.212	0.861719950873643\\
0.216	0.851149670967633\\
0.22	0.840539984282721\\
0.224	0.829939036552137\\
0.228	0.819394933852578\\
0.232	0.808955524303485\\
0.236	0.798668180936894\\
0.24	0.788579586723195\\
0.244	0.778735522728323\\
0.248	0.769180660363702\\
0.252	0.759958358671698\\
0.256	0.751110467566469\\
0.26	0.742677137923102\\
0.264	0.734696639376818\\
0.268	0.727205186659066\\
0.272	0.720236775258568\\
0.276	0.713823027153077\\
0.28	0.707993047311898\\
0.284	0.702773291620344\\
0.288	0.698187446825491\\
0.292	0.694256323048002\\
0.296	0.690997759347815\\
0.3	0.688426542772209\\
0.304	0.68655434125361\\
0.308	0.685389650661647\\
0.312	0.684937756249711\\
0.316	0.684931506849315\\
0.32	0.684931506849315\\
0.324	0.684931506849315\\
0.328	0.684931506849315\\
0.332	0.684931506849315\\
0.336	0.684931506849315\\
0.34	0.684931506849315\\
0.344	0.684931506849315\\
0.348	0.684931506849315\\
0.352	0.684931506849315\\
0.356	0.684931506849315\\
0.36	0.684931506849315\\
0.364	0.684931506849315\\
0.368	0.684931506849315\\
0.372	0.684931506849315\\
0.376	0.684931506849315\\
0.38	0.684931506849315\\
0.384	0.684931506849315\\
0.388	0.684931506849315\\
0.392	0.684931506849315\\
0.396	0.684931506849315\\
0.4	0.684931506849315\\
0.404	0.684931506849315\\
0.408	0.684931506849315\\
0.412	0.684931506849315\\
0.416	0.684931506849315\\
0.42	0.684931506849315\\
0.424	0.684931506849315\\
0.428	0.684931506849315\\
0.432	0.684931506849315\\
0.436	0.684931506849315\\
0.44	0.684931506849315\\
0.444	0.684931506849315\\
0.448	0.684931506849315\\
0.452	0.684931506849315\\
0.456	0.684931506849315\\
0.46	0.684931506849315\\
0.464	0.684931506849315\\
0.468	0.684931506849315\\
0.472	0.684931506849315\\
0.476	0.684931506849315\\
0.48	0.684931506849315\\
0.484	0.684931506849315\\
0.488	0.684931506849315\\
0.492	0.684931506849315\\
0.496	0.684931506849315\\
0.5	0.684931506849315\\
0.504	0.684931506849315\\
0.508	0.684931506849315\\
0.512	0.684931506849315\\
0.516	0.684931506849315\\
0.52	0.684931506849315\\
0.524	0.684931506849315\\
0.528	0.684931506849315\\
0.532	0.684931506849315\\
0.536	0.684931506849315\\
0.54	0.684931506849315\\
0.544	0.684931506849315\\
0.548	0.684931506849315\\
0.552	0.684931506849315\\
0.556	0.684931506849315\\
0.56	0.684931506849315\\
0.564	0.684931506849315\\
0.568	0.684931506849315\\
0.572	0.684931506849315\\
0.576	0.684931506849315\\
0.58	0.684931506849315\\
0.584	0.684931506849315\\
0.588	0.684931506849315\\
0.592	0.684931506849315\\
0.596	0.684931506849315\\
0.6	0.684931506849315\\
0.604	0.684931506849315\\
0.608	0.684931506849315\\
0.612	0.684931506849315\\
0.616	0.684931506849315\\
0.62	0.684931506849315\\
0.624	0.684931506849315\\
0.628	0.684931506849315\\
0.632	0.684931506849315\\
0.636	0.684931506849315\\
0.64	0.684931506849315\\
0.644	0.684931506849315\\
0.648	0.684931506849315\\
0.652	0.684931506849315\\
0.656	0.684931506849315\\
0.66	0.684931506849315\\
0.664	0.684931506849315\\
0.668	0.684931506849315\\
0.672	0.684931506849315\\
0.676	0.684931506849315\\
0.68	0.684931506849315\\
0.684	0.684931506849315\\
0.688	0.684931506849315\\
0.692	0.684931506849315\\
0.696	0.684931506849315\\
0.7	0.684931506849315\\
0.704	0.684931506849315\\
0.708	0.684931506849315\\
0.712	0.684931506849315\\
0.716	0.684931506849315\\
0.72	0.684931506849315\\
0.724	0.684931506849315\\
0.728	0.684931506849315\\
0.732	0.684931506849315\\
0.736	0.684931506849315\\
0.74	0.684931506849315\\
0.744	0.684931506849315\\
0.748	0.684931506849315\\
0.752	0.684931506849315\\
0.756	0.684931506849315\\
0.76	0.684931506849315\\
0.764	0.684931506849315\\
0.768	0.684931506849315\\
0.772	0.684931506849315\\
0.776	0.684931506849315\\
0.78	0.684931506849315\\
0.784	0.684931506849315\\
0.788	0.684931506849315\\
0.792	0.684931506849315\\
0.796	0.684931506849315\\
0.8	0.684931506849315\\
0.804	0.684931506849315\\
0.808	0.684931506849315\\
0.812	0.684931506849315\\
0.816	0.684931506849315\\
0.82	0.684931506849315\\
0.824	0.684931506849315\\
0.828	0.684931506849315\\
0.832	0.684931506849315\\
0.836	0.684931506849315\\
0.84	0.684931506849315\\
0.844	0.684931506849315\\
0.848	0.684931506849315\\
0.852	0.684931506849315\\
0.856	0.684931506849315\\
0.86	0.684931506849315\\
0.864	0.684931506849315\\
0.868	0.684931506849315\\
0.872	0.684931506849315\\
0.876	0.684931506849315\\
0.88	0.684931506849315\\
0.884	0.684931506849315\\
0.888	0.684931506849315\\
0.892	0.684931506849315\\
0.896	0.684931506849315\\
0.9	0.684931506849315\\
0.904	0.684931506849315\\
0.908	0.684931506849315\\
0.912	0.684931506849315\\
0.916	0.684931506849315\\
0.92	0.684931506849315\\
0.924	0.684931506849315\\
0.928	0.684931506849315\\
0.932	0.684931506849315\\
0.936	0.684931506849315\\
0.94	0.684931506849315\\
0.944	0.684931506849315\\
0.948	0.684931506849315\\
0.952	0.684931506849315\\
0.956	0.684931506849315\\
0.96	0.684931506849315\\
0.964	0.684931506849315\\
0.968	0.684931506849315\\
0.972	0.684931506849315\\
0.976	0.684931506849315\\
0.98	0.684931506849315\\
0.984	0.684931506849315\\
0.988	0.684931506849315\\
0.992	0.684931506849315\\
0.996	0.684931506849315\\
1	0.684931506849315\\
};
\draw[color=black] (0.8, 0.1) node {$\theta>\theta_\text{max}$}; 
\end{axis}
\end{tikzpicture}%
		\end{subfigure}	
		\caption{New analytical model: Normalized tangential traction $\sigzt/\ttmax$ as a function of the radius $r/R$ of the contact surface (see Eq.~\eqref{eq:new_ana}) at different rotation angles.}
		\label{img:ana_traction}
	\end{figure}

	The tangential contact traction as a function of the rotation angle can then be expressed as
	\begin{equation} \label{eq:new_ana}
		\displaystyle{
			\sigzt(\theta,r) = \begin{cases}
				\ttmax\frac{r}{c(\theta)}, & \text{for } r < c \text{ (sticking)}, \\
				\mu p, & \text{for } r \ge c \text{ (sliding)},
			\end{cases}} 
	\end{equation}
	with $\mu = \mu(\gs)$ from Eq.~\eqref{eq:mod_mu}. \fig{img:ana_traction} illustrates the variation of the tangential contact traction~\eqref{eq:new_ana} for $\theta = \thetalin$, $\theta = \thetamax$, and some $\theta$ larger than $\thetamax$.
	
	While the two analytical models presented in \sect{s:case_ana} and \sect{s:case_ana_new} both assume circular crack propagation from outside to inside and determine the stick/slip transition by a critical (or crack) radius, the analytical model from Ref. \cite{mathieu2012} imposes the crack radius and computes the corresponding rotation angle, while the new model directly imposes the rotation angle. 
	%--------------------------------------------------------------------
	%--------------------------------------------------------------------
	%
	%--------------------------------------------------------------------
	% Numerical setup
	%--------------------------------------------------------------------
	\subsection{Numerical setup} \label{s:case_setup}
	As in the experiments, we consider a coin-shaped cylindrical implant with dimensions $R=\SI{2.5}{\milli\m}$ and $H=\SI{3}{\milli\m}$.
	The bone sample is modeled as a rectangular cuboid with dimensions 12.5 $\times$ 12.5 $\times$ \SI{5}{\milli\m}. 
	The implant is positioned at the center of the upper bone surface.
	The bodies are meshed according to the parameters given in \tab{tab:mesh}, where $n_\text{e}$ denotes the number of elements of the body/surface and $n_\text{gp}$ denotes the number of Gauss-points per element.
	While the bulk is discretized with linear Lagrangian shape functions, the contact surfaces are discretized with non--uniform rational B-Splines (NURBS) (see Appendix \ref{s:appndx_fem}).
	The finite element mesh is pictured in \fig{img:exp2}. 
	To justify this coarse discretization, a refinement analysis of the mesh and the load step size is presented in Appendix \ref{s:appndx_conv}. 
	 
	Contact is computed with a penalty regularization, and the corresponding penalty parameter is set according to the Young's modulus of bone to $\epsilon= \Eb/L_0$, with $L_0=\SI{0.01}{m}$.
	
	The lower surface of the bone block is fixed in all directions, while the sides of the bone block are fixed in their corresponding normal direction. Our modified Coulomb's friction model requires a contact pressure during sliding, see Eq.~\eqref{eq:coulomb} and \eqref{eq:new_ana}. We generate this pressure by applying a uniform vertical displacement $d$ at the upper surface of the implant. 
	Then, the implant is rotated by \ang{10} around its central axis with a constant load step size of \ang{0.01}.
	
	All computations in Secs.~\ref{s:res_param}-\ref{s:res_work} use homogeneous initial bonding ($\phio(\bx)=1, \, \forall \bx \in \csurf$), while \sect{s:res_osseo} presents cases with inhomogeneous initial bonding.
	%--------------------------------------------------------------------
	\begin{table}[H]	
		\centering
		\begin{tabular}{|c||c|c|c|}
			\hline 
			body & $n_\text{e}$ & type of shape fcts. & $n_\text{gp}$\\ 
			\hline 
			\hline 
			implant bulk & 18 & linear Lagrange & $2\times 2\times 2$ \\
			bone bulk & 450 & linear Lagrange & $2\times 2\times 2$ \\
			lower implant surface & 9 & quadratic NURBS & $5\times 5$ \\
			upper bone surface & 225 & quadratic NURBS & $5\times 5$\\
			\hline
		\end{tabular} 
		\caption{Parameters of the finite element mesh: Number of elements $n_\text{e}$, type of shape functions and number of Gauss-points per element $n_\text{gp}$ for the two bodies and their contact surfaces.}
		\label{tab:mesh}
	\end{table}
	%--------------------------------------------------------------------
	For both bodies, the Neo-Hookean material model of Eq.~\eqref{eq:neohooke} is used.
	The material properties for the implant are those of titanium alloy (Ti-6Al-4V: $E_i=\SI{113}{\giga\Pa}$, $\nu_i=0.3$).
	The material properties of the bone and the friction parameters $\mus,\,\muk,\,\as,\,\bs,$ and the contact pressure $p$ have to be determined by a parameter study and are presented in \sect{s:res_param}.
	All simulations were performed with an in-house, MATLAB-based solver\footnote{(R2018b, The MathWorks, Natick, MA, USA)}.
%
%--------------------------------------------------------------------
% Results
%--------------------------------------------------------------------
\section{Results} \label{s:results}
In the following, the results obtained with the new analytical model and numerical study are presented and compared to the experimental and analytical results corresponding to the reference study. First, the parameter estimation and the subsequent error estimation for homogeneous osseointegration are presented. Second, the torque-per-angle curves of the different models are compared and the debonding behavior of the implant is discussed. Third, the work of adhesion and frictional energy loss of the models are compared. Last, several cases of partial osseointegration are presented and compared with the homogeneous case.
	%
	%--------------------------------------------------------------------
	% Parameter estimation
	%--------------------------------------------------------------------
	\subsection{Parameter calibration} \label{s:res_param}
	During the parameter estimation stage, the Poisson ratio of bone is fixed to $\nu_\text{b} = 0.3$. The remaining parameters $\Gb,\,d,\,\mus,\,\as,\,\bs$ are determined by minimizing the mean relative error
	\begin{equation}
		e^\mathrm{mp}_{T} = \underset{\theta\in\left[\ang{0},\ang{10}\right]}{\mathrm{mean}} \left( \lv \frac{\Texp(\theta)-T(\theta)}{\Texp(\theta)} \rv \right) \label{eq:mpe},
	\end{equation}
	where $\Texp$ is defined as the torque over the rotation angle $\theta$ obtained from the corresponding experiment. The shear modulus $\Gb$ is calibrated using the initial slope of the linear part of the torque-per-angle curve (i.e. $T(\theta\le\thetalin)$) (e.g. see \fig{img:torque_a}).
	The other parameters depend on the friction coefficient $\muk$, which is investigated at the fixed values $\muk=[0.2, 0.25, 0.3, 0.35, 0.4, 0.45, 0.5]$, which corresponds to the typical range found in the literature~\cite{biemond2011,grant2007,rancourt1990,shirazi-adl1993}. The vertical displacement $d$ is calibrated from steady state sliding, which is considered here to occur for $\theta > \ang{9}$.
	We thus minimize Eq.~\eqref{eq:mpe} for $\theta\in[\ang{9},\ang{10}]$ to calibrate $d$. 	
	There, the implant is assumed to be fully debonded and thus, tangential sliding contact is governed by Coulomb's law~\eqref{eq:coulomb} with $\mu=\muk$.
	 Once $d$ is determined, it is considered constant for all rotation angles $\theta$. 
	
	The initial friction coefficient $\mus$ is calibrated using $\Tmax$ (as $\ttmax=\mus\,p$) (see \tab{tab:datasets}). 
	Finally, the parameters $\as$ and $\bs$ are then determined by minimizing Eq.~\eqref{eq:mpe} for the whole torque-per-angle curve.
	
	The bone shear modulus can be estimated from the new analytical model. 
	Due to the non-homogeneous pressure distribution, the displacement $d$ and subsequently all other parameters are determined using the numerical model. 
	However, adjusting $d$ for the new analytical model of \sect{s:case_ana_new} leads to the same estimated values for $\mus, \as,$ and $\bs$. 
	%--------------------------------------------------------------------
	\begin{table}[htbp]	
		\centering
		\begin{tabular}{|c||c|c|c|c|}
			\hline 
			data set & $\Tmax$ [\si{\N\m}] & $\Tinf$ [\si{\N\m}] & $\thetalin$ & $\thetamax$ \\ 
			\hline 
			\hline 
			1 & 0.0538 & 0.0368 & \ang{0.13} & \ang{1.13} \\
			2 & 0.0595 & 0.0444 & \ang{0.11} & \ang{2.02} \\
			\hline
		\end{tabular} 
		\caption{Data sets used for the parameter estimation. Data set 1 and 2 correspond to the data shown in Figure 4a and 4b of Mathieu~\etal~\cite{mathieu2012}, respectively. $\thetalin$ denotes the limit of the elastic part of the deformation, while $\thetamax$ is the location of the maximum torque $\Tmax$. $\Tinf$ denotes the torque for a fully debonded implant and is taken at \ang{10}.}
		\label{tab:datasets}
	\end{table}
	%--------------------------------------------------------------------
	
	\tab{tab:param_estim1} shows the estimated shear moduli and the corresponding Young's moduli obtained from the numerical parameter estimation for the two data sets, compared to the computed shear moduli obtained in Ref.~\cite{mathieu2012}. 
	The estimated shear moduli of 7 and 8~\si{\giga\Pa} are higher than the reported values of 2--6~\si{\giga\Pa}~\cite{sharma2012,tang2015}), while the corresponding Young's moduli of 18 and \SI{21}{\giga\Pa} are in good agreement with experimental data from the literature~\cite{novitskaya2011,rho1993} and previous studies~\cite{vayron2011,vayron2012}. In Ref.~\cite{mathieu2012} the model was fit to match the peak and the decrease in torsion, which results in a considerable error for the shear modulus, as shown in \tab{tab:param_estim1} and \figs{img:torque}. As our proposed model allows for more control, it is possible to match more characteristics of the torque curve, such as the initial slope, the peak, the softening and the torque for complete debonding.

	\tab{tab:param_estim2} shows the results of the numerical parameter estimation of the different data sets and the corresponding mean percentage error, for the chosen levels of friction coefficient $\muk$. 
	For the analyzed $\muk$, the corresponding initial friction coefficient $\mus$ lies between 0.29 and 0.73 and agrees well with the values of 0.28--1.1 reported in the literature~\cite{biemond2011,damm2015,grant2007,rancourt1990,shirazi-adl1993,zhang1999}.
	%--------------------------------------------------------------------
	\begin{table}[htbp]
		\centering
		\begin{tabular}{|c||c|c|c|}
			\hline 
			data set  & model & $\Gb$ [\si{\giga\Pa}] & $\Eb$ [\si{\giga\Pa}] \\ 
			\hline \hline 
			 & literature \cite{novitskaya2011,rho1993,sharma2012,tang2015,vayron2011,vayron2012} & 2--6 & 15--23 \\
			\hline \hline 
			\multirow{2}{*}{1} & analytical solution from Ref. \cite{mathieu2012} & 0.04 & -- \\
			& present numerical solution & 7 & 18 \\
			\hline
			\hline
			\multirow{2}{*}{2} & analytical solution from Ref. \cite{mathieu2012} & 0.02 & -- \\
			& present numerical solution & 8 & 21 \\
			\hline
		\end{tabular} 
		\caption{Results for the parameter estimation of the two data sets for values independent of the friction coefficient $\muk$. Shown are results from the analytical model presented in \cite{mathieu2012} (see Eqs.~\eqref{eq:ana_old1},\eqref{eq:ana_old2}) and the present numerical solution. The estimated parameter is the bone shear modulus $\Gb$. The Young's modulus $\Eb$ then follows from Eq.~\eqref{eq:young}.}
		\label{tab:param_estim1}
	\end{table}
	%--------------------------------------------------------------------
		Increasing $\muk$ results in a smaller normal force and a higher $\mus$. Calibrating the parameters using $\Tmax$ and $\Tinf$ results in almost identical curves for all tested values of $\muk$. 
	Therefore, for the second data set only $\muk = 0.3$~\cite{rancourt1990} and $\muk = 0.4$~\cite{grant2007,shirazi-adl1993} are investigated, which are the values most commonly reported for the interface between cortical bone and polished metal implants.
	
	As shown in \tab{tab:param_estim2}, the parameters $\as$ and $\bs$ only depend on the shear modulus and on the overall shape of the torque curve, i.e., the width of the peak and the slope of $T(\theta)$. 
	The computed values for $\as$ indicate that adhesion takes effect for a sliding distance up to 22 and \SI{26}{\micro\m}, respectively. When these values are exceeded, the implant starts to debond, which is indicated by a decreasing friction coefficient (see \fig{img:mu_gs}). 
	This observation is in accordance with the reported threshold for micro-motion of the BII, where no deformation occurs. 
	In most studies, a value of up to \SI{50}{\micro\m} is reported~\cite{bragdon1996,fitzpatrick2014}, while values exceeding \SI{150}{\micro\m} have shown to inhibit bone growth and promote bone loss~\cite{jasty1997,pilliar1986}. However, these values were only reported for normal displacement and may vary for tangential displacement.
	%
	%--------------------------------------------------------------------
	\begin{table}[H]	
		\centering
		\begin{tabular}{|c||c|c|c|c|c|c|c|}
			\hline 
			data set &  $\muk$ & $d$ [\si{\micro\m}] & $\bar{p}$ [\si{\mega\Pa}] & $\mus$ & $\as$ [\si{\micro\m}] & $\bs$ & $\errmp$ [\si{\percent}] \\ 
			\hline \hline 
			\multirow{8}{*}{1} & ana.~\cite{mathieu2012} & -- & -- & -- & -- & -- &  7.7200 \\
			\cline{2-8}
			&  0.20 & 9.8 & 5.2 & 0.29 & \multirow{7}{*}{22 } & \multirow{7}{*}{0.74} & 2.2390\\
			&  0.25 & 7.8 & 4.2 & 0.37 & & & 2.2395 \\
			&  0.30 & 6.5 & 3.5 & 0.44 & & & 2.2397 \\ 
			&  0.35 & 5.6 & 3.0 & 0.51 & & & 2.2399 \\
			&  0.40 & 4.9 & 2.6 & 0.58 & & & 2.2400 \\ 
			&  0.45 & 4.4 & 2.3 & 0.66 & & & 2.2401 \\
			&  0.50 & 3.9 & 2.1 & 0.73 & & & 2.2401 \\ 
			\hline \hline
			\multirow{3}{*}{2} & ana.~\cite{mathieu2012} & -- & -- & -- & -- & -- & 11.6231\\
			\cline{2-8}
			&  0.30 & 6.8 & 4.1 & 0.41  & \multirow{2}{*}{26 } & \multirow{2}{*}{1.86} & 2.1499 \\ 	
			&  0.40 & 5.1 & 3.1 & 0.55 & & & 2.1500 \\ 
			\hline
		\end{tabular} 
		\caption{Results for the parameter estimation of the two data sets. Shown are results from the analytical model presented in Ref.~\cite{mathieu2012} (see Eqs.~\eqref{eq:ana_old1},\eqref{eq:ana_old2}) and the present numerical solution. The estimated parameters are the enforced normal displacement $d$, corresponding average contact pressure $\bar{p}$, friction coefficient for the unbroken state $\mus$, sliding threshold $\as$, and transition factor $\bs$.}
		\label{tab:param_estim2}
	\end{table}
	%--------------------------------------------------------------------

	%--------------------------------------------------------------------
	% Torque
	%--------------------------------------------------------------------
	\subsection{Torque curves and debonding behavior}\label{s:res_torque}
\begin{figure}[H]
		\centering
		\begin{subfigure}[t]{\textwidth}
			\centering
			\input{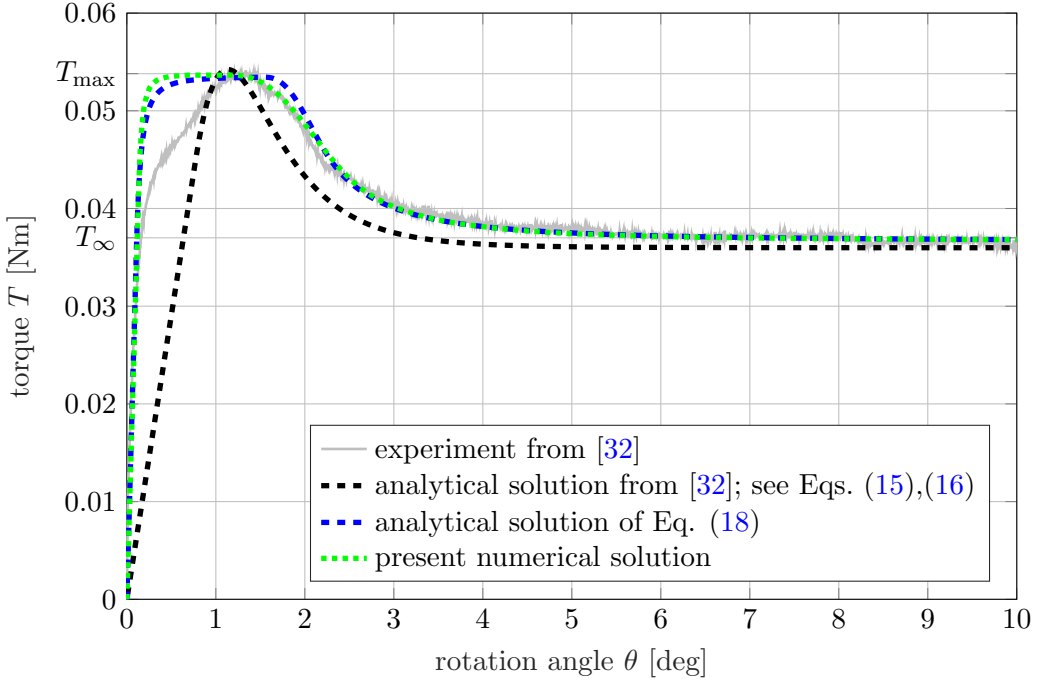}
			\caption{First data set.}
			\label{img:torque_a}
		\end{subfigure}
		\hspace{5mm}
		\begin{subfigure}[t]{\textwidth}
			\centering
			\input{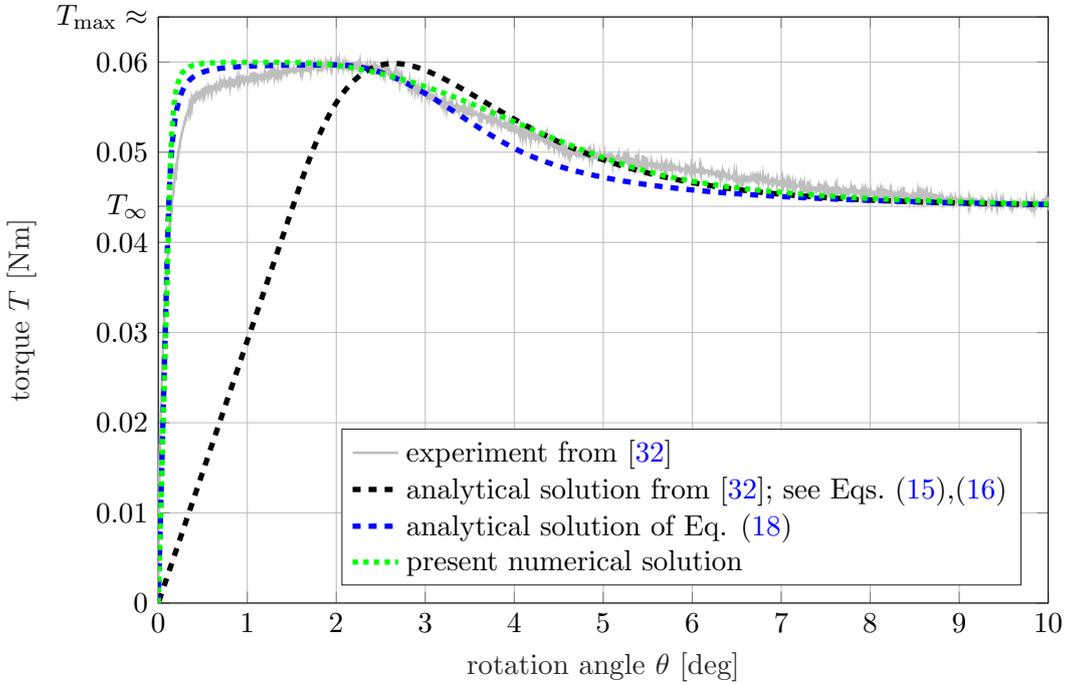}
			\caption{Second data set.}
			\label{img:torque_b}
		\end{subfigure}
		\caption{Variation of the Torque $T$ as a function of the imposed rotation angle~$\theta$.}
		\label{img:torque}
	\end{figure}

	The corresponding curves representing the variation of the torque as a function of the angle of rotation are shown in \figs{img:torque_a} and \ref{img:torque_b}. 
	The resulting torque obtained with the new analytical model described in \sect{s:case_ana_new} and the numerical solution of the proposed friction model of \sect{s:model_modcoulomb} (using $\muk = 0.4$) are very close. 
	While the analytical solution shows a slightly closer fit to the experimental data before the peak in torque (resulting in an error of 2.18 \si{\percent} for the first data set and 2.83 \si{\percent} for the second data set), the numerical solution provides a better estimation of the behavior after debonding. 
		
	The errors obtained with the numerical solution are given in \tab{tab:param_estim2} and are compared to the analytical model from the reference study. 
	Overall, the numerical solutions yield the best agreement with the experimental data, especially concerning the initial slope (i.e. stiffness) of the torque and the decrease after its peak. 
	The torque curves show a flat plateau at the peak, which comes from the behavior induced by the sliding threshold $\as$. 
	Increasing $\as$ induces an offset of the debonding process and thus, results in an elongated peak.
	Another difference is shown in \fig{img:torque_b} for $\theta > \ang{2.5}$, where the decrease of the torque is not exactly reproduced. 
	A different transition function $\phi$ may allow for a closer fit there.
	
\begin{figure}[htbp]
		\centering
		\begin{overpic}[width=0.146\linewidth]{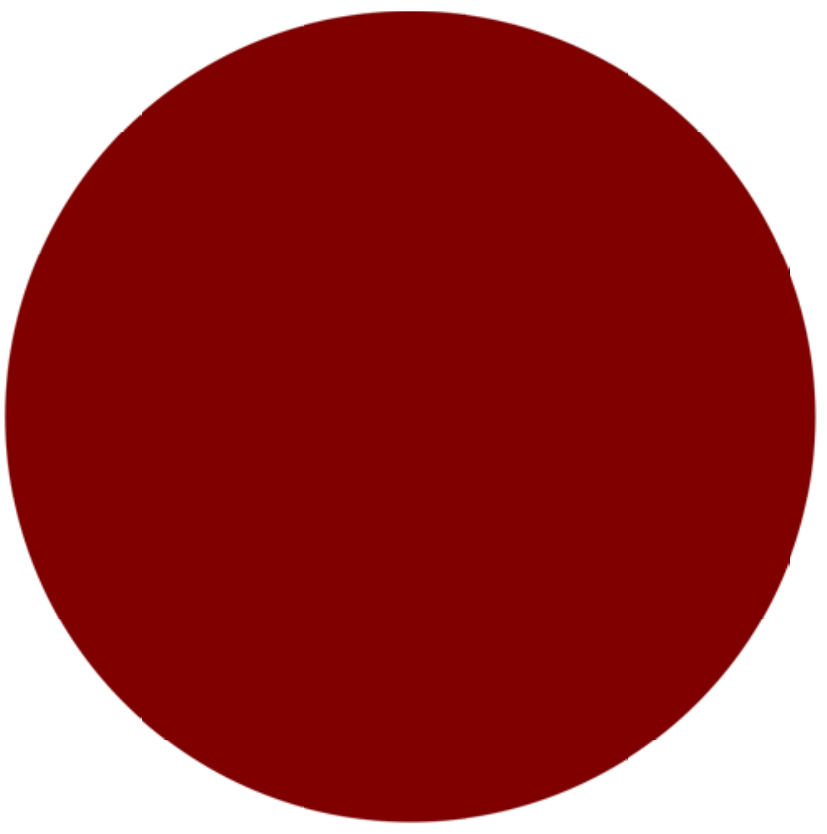}
			\put (84,5) {\small$\ang{1}$}
		\end{overpic}
		\begin{overpic}[width=0.146\linewidth]{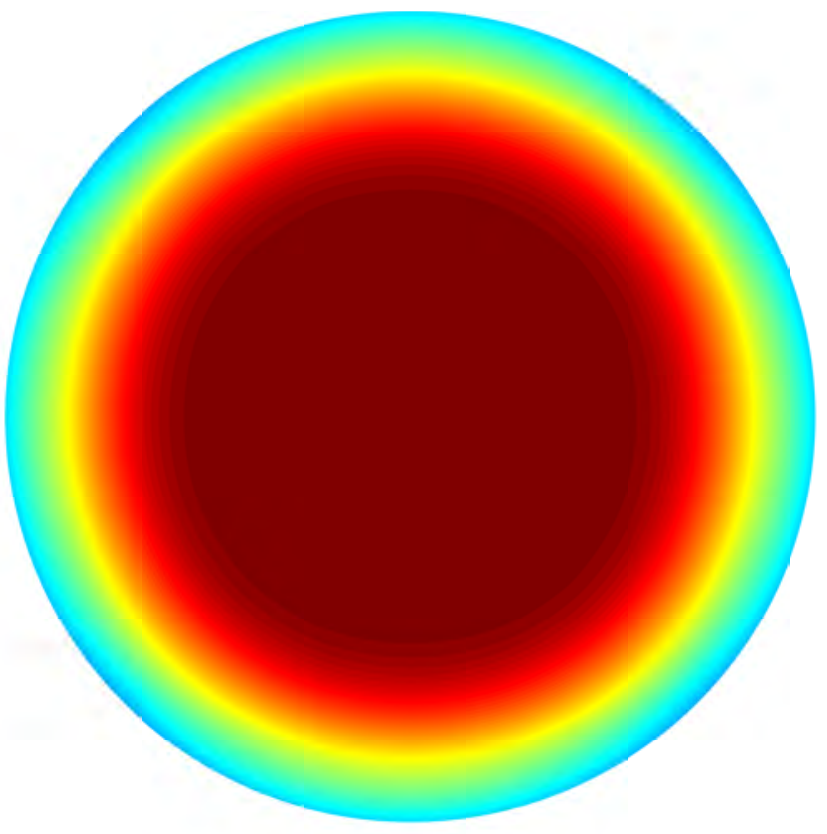}
			\put (84,5) {\small$\ang{2}$}
		\end{overpic}
		\begin{overpic}[width=0.146\linewidth]{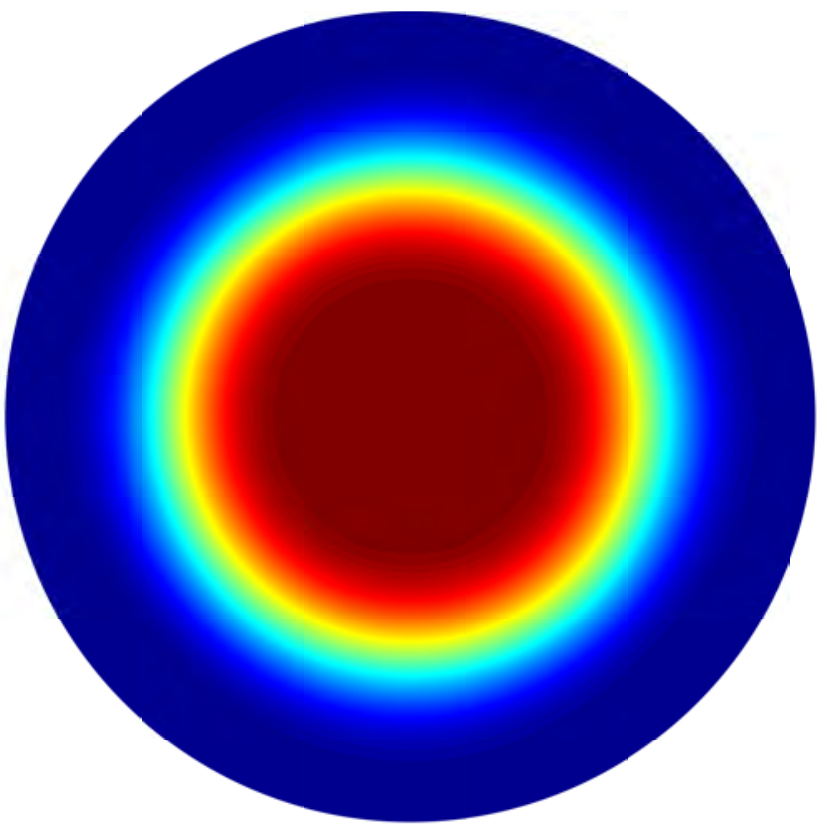}
			\put (84,5) {\small$\ang{3}$}
		\end{overpic}
		\begin{overpic}[width=0.146\linewidth]{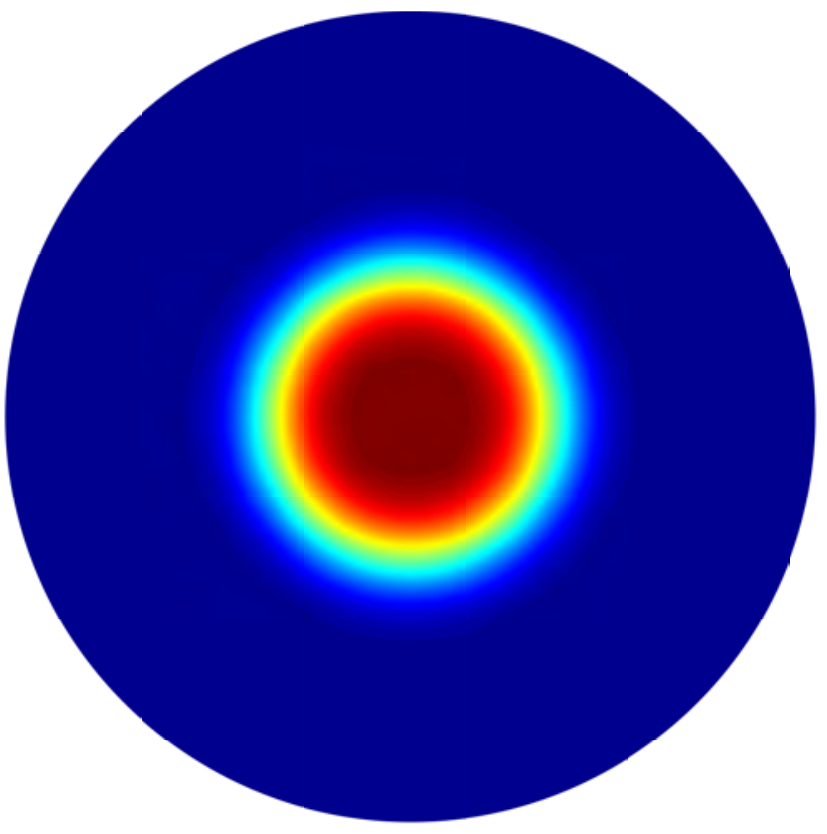}
			\put (84,5) {\small$\ang{5}$}
		\end{overpic}
		\begin{overpic}[width=0.146\linewidth]{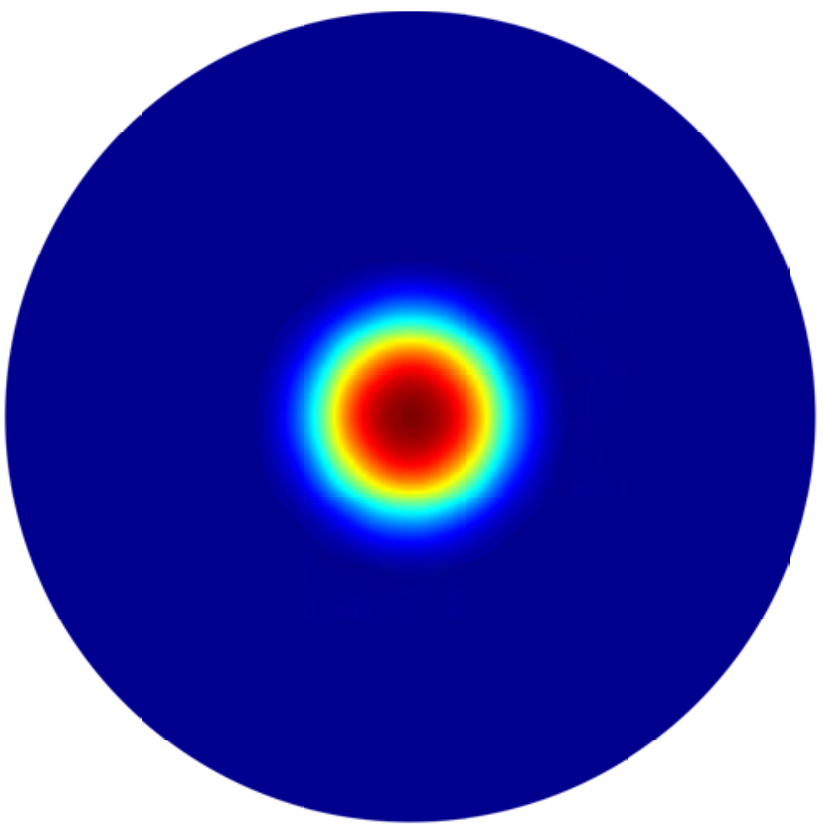}
			\put (84,5) {\small$\ang{8}$}
		\end{overpic}
		\begin{overpic}[width=0.146\linewidth]{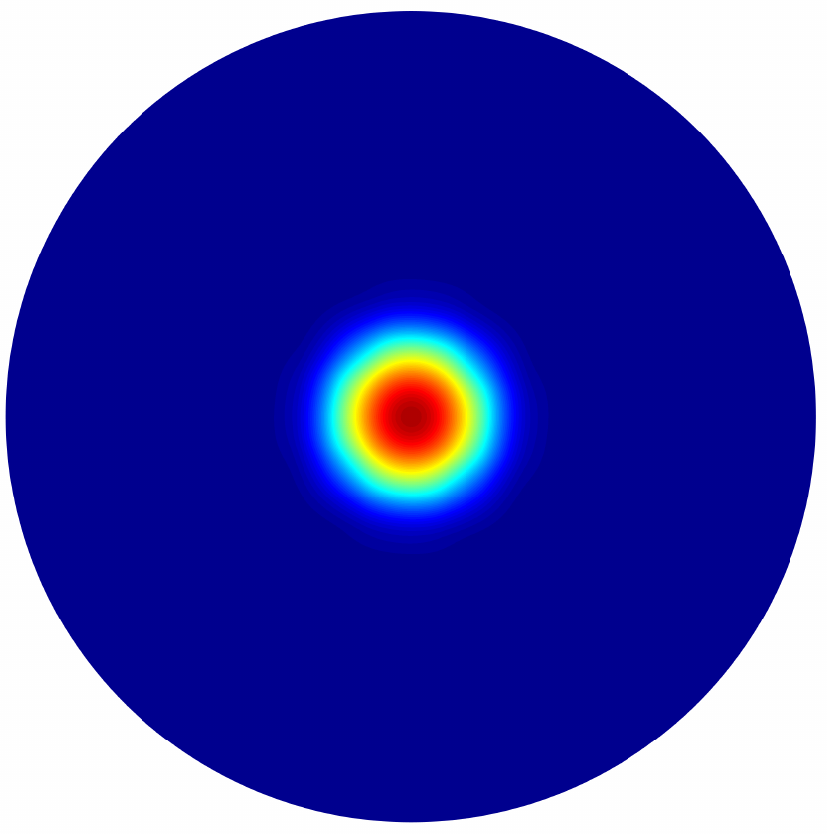}
			\put (84,5) {\small$\ang{10}$}
		\end{overpic}
		\hspace{0.5mm}
		\begin{subfigure}{0.05\linewidth}
			\vspace{-20mm}
			\def\svgwidth{0.4\textwidth}
%% Accompanies image file 'colorbar_mufs.pdf' (pdf, eps, ps)
\begingroup%
  \makeatletter%
  \providecommand\color[2][]{%
    \errmessage{(Inkscape) Color is used for the text in Inkscape, but the package 'color.sty' is not loaded}%
    \renewcommand\color[2][]{}%
  }%
  \providecommand\transparent[1]{%
    \errmessage{(Inkscape) Transparency is used (non-zero) for the text in Inkscape, but the package 'transparent.sty' is not loaded}%
    \renewcommand\transparent[1]{}%
  }%
  \providecommand\rotatebox[2]{#2}%
  \newcommand*\fsize{\dimexpr\f@size pt\relax}%
  \newcommand*\lineheight[1]{\fontsize{\fsize}{#1\fsize}\selectfont}%
  \ifx\svgwidth\undefined%
    \setlength{\unitlength}{44.31712735bp}%
    \ifx\svgscale\undefined%
      \relax%
    \else%
      \setlength{\unitlength}{\unitlength * \real{\svgscale}}%
    \fi%
  \else%
    \setlength{\unitlength}{\svgwidth}%
  \fi%
  \global\let\svgwidth\undefined%
  \global\let\svgscale\undefined%
  \makeatother%
  \begin{picture}(1,5.89808633)%
    \lineheight{1}%
    \setlength\tabcolsep{0pt}%
    \put(0,0.1){\includegraphics[width=\unitlength,page=1]{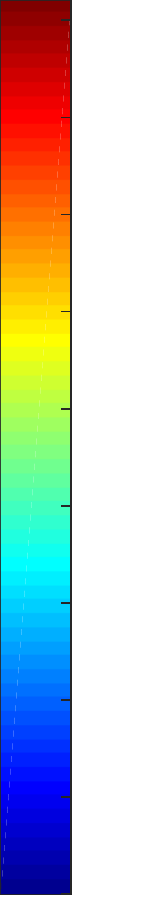}}%
    \put(0.0,6.5){\makebox(0,0)[lt]{\lineheight{1.25}\smash{\begin{tabular}[t]{l}\small{$\mu$}\end{tabular}}}}%
    \put(0.62,5.6){\makebox(0,0)[lt]{\lineheight{1.25}\smash{\begin{tabular}[t]{l}\footnotesize{0.58}\end{tabular}}}}%
    \put(0.62,2.8){\makebox(0,0)[lt]{\lineheight{1.25}\smash{\begin{tabular}[t]{l}\footnotesize{0.5}\end{tabular}}}}%
    \put(0.62,0.1){\makebox(0,0)[lt]{\lineheight{1.25}\smash{\begin{tabular}[t]{l}\footnotesize{0.4}\end{tabular}}}}%
  \end{picture}%
\endgroup%

		\end{subfigure}
		\begin{overpic}[width=0.146\linewidth]{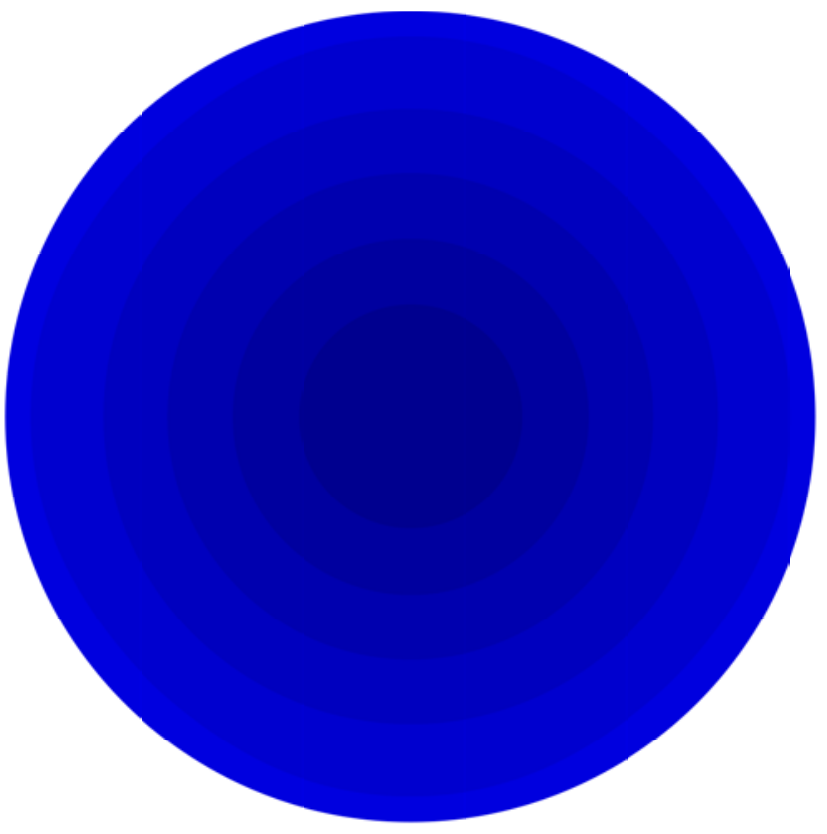}
			\put (84,5) {\small$\ang{1}$}
		\end{overpic}
		\begin{overpic}[width=0.146\linewidth]{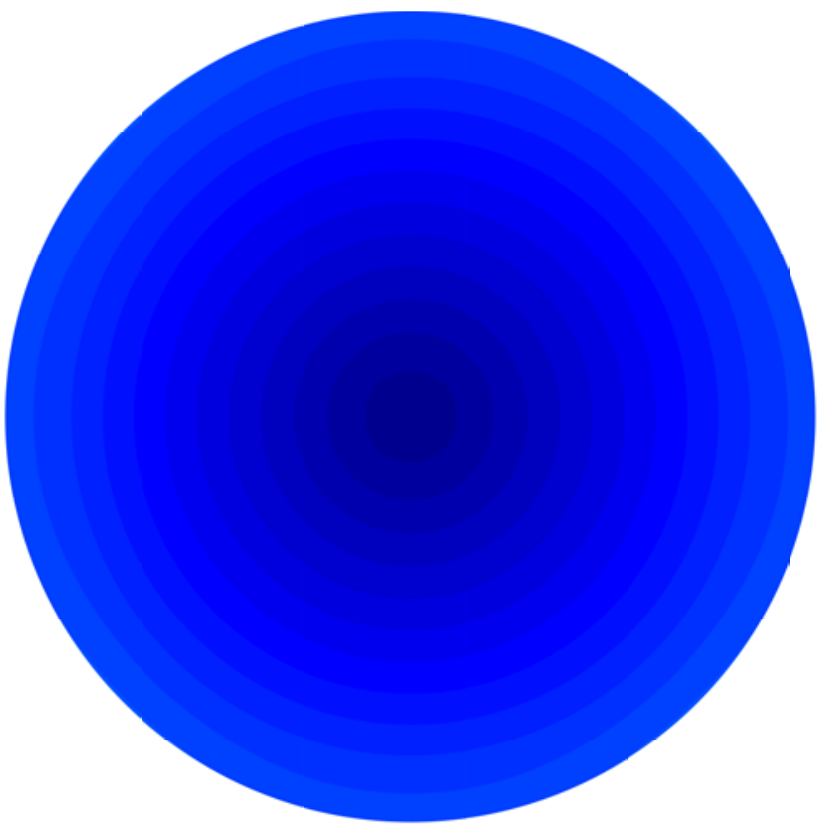} 		
			\put (84,5) {\small$\ang{2}$}
		\end{overpic}
		\begin{overpic}[width=0.146\linewidth]{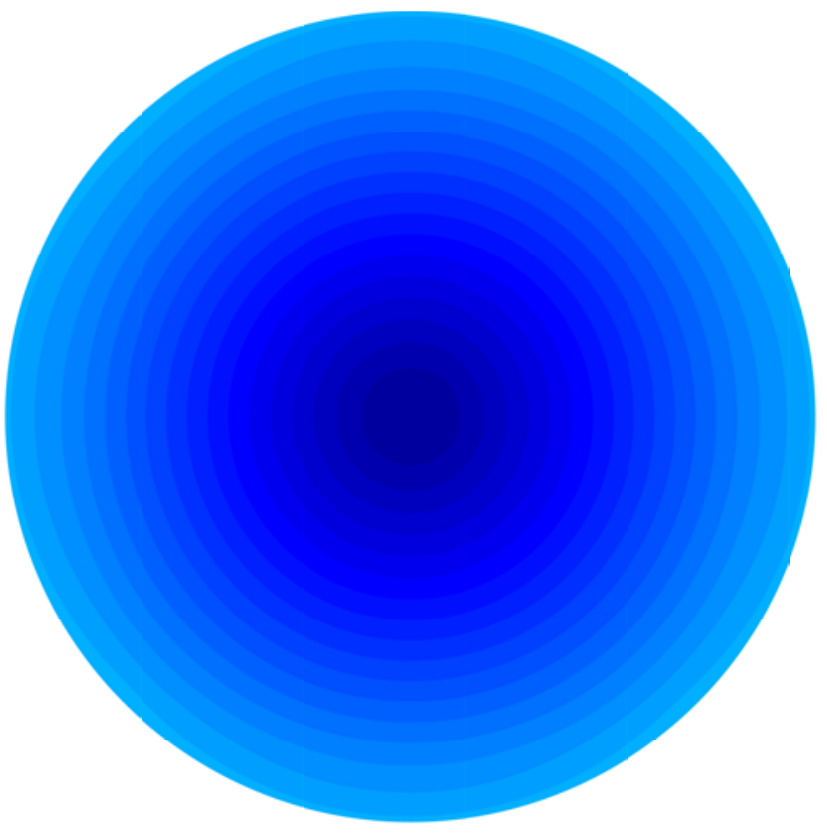} 	
			\put (84,5) {\small$\ang{3}$}
		\end{overpic}
		\begin{overpic}[width=0.146\linewidth]{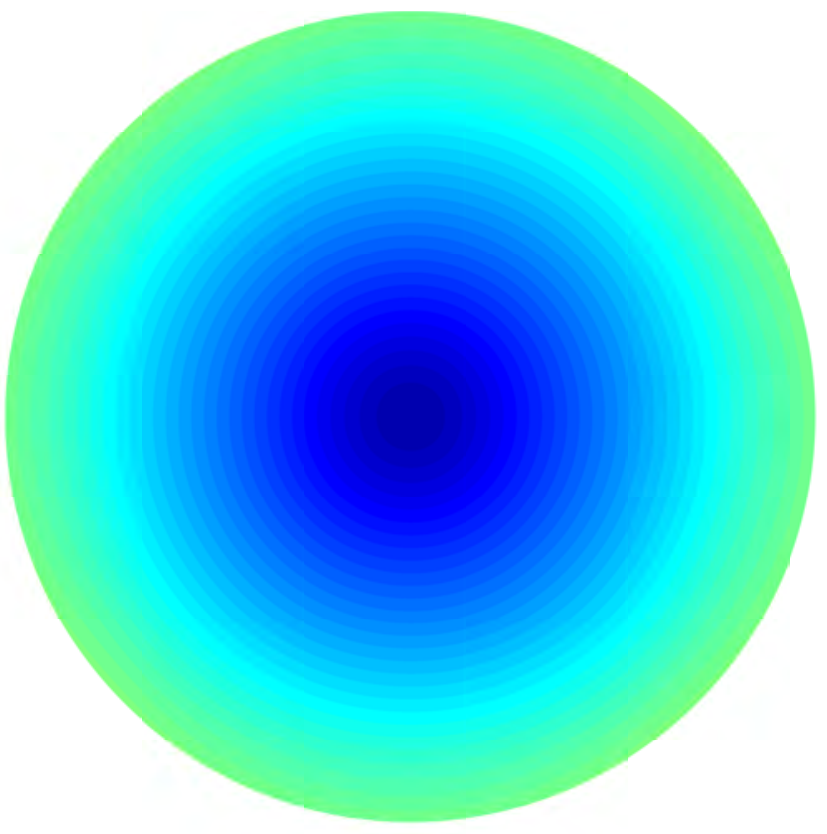} 		
			\put (84,5) {\small$\ang{5}$}
		\end{overpic}
		\begin{overpic}[width=0.146\linewidth]{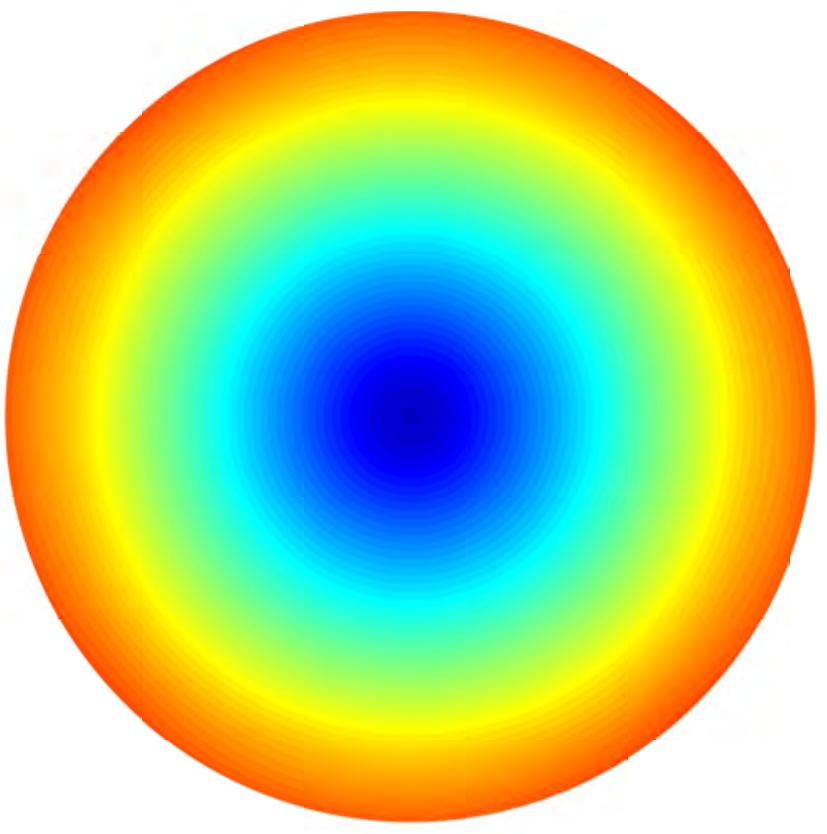} 		
			\put (84,5) {\small$\ang{8}$}
		\end{overpic}
		\begin{overpic}[width=0.146\linewidth]{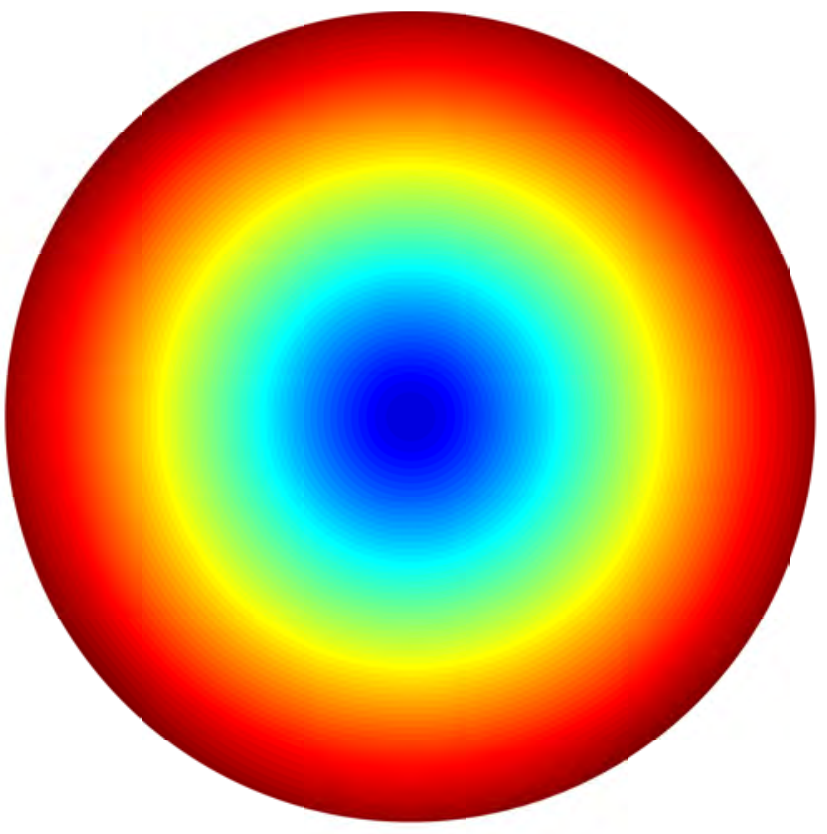} 		
			\put (84,5) {\small$\ang{10}$}
		\end{overpic}
		\hspace{0.5mm}
		\begin{subfigure}{0.05\linewidth}
			\vspace{-20mm}
			\def\svgwidth{0.4\textwidth}
%% Accompanies image file 'colorbar_mufs.pdf' (pdf, eps, ps)
\begingroup%
  \makeatletter%
  \providecommand\color[2][]{%
    \errmessage{(Inkscape) Color is used for the text in Inkscape, but the package 'color.sty' is not loaded}%
    \renewcommand\color[2][]{}%
  }%
  \providecommand\transparent[1]{%
    \errmessage{(Inkscape) Transparency is used (non-zero) for the text in Inkscape, but the package 'transparent.sty' is not loaded}%
    \renewcommand\transparent[1]{}%
  }%
  \providecommand\rotatebox[2]{#2}%
  \newcommand*\fsize{\dimexpr\f@size pt\relax}%
  \newcommand*\lineheight[1]{\fontsize{\fsize}{#1\fsize}\selectfont}%
  \ifx\svgwidth\undefined%
    \setlength{\unitlength}{44.31712735bp}%
    \ifx\svgscale\undefined%
      \relax%
    \else%
      \setlength{\unitlength}{\unitlength * \real{\svgscale}}%
    \fi%
  \else%
    \setlength{\unitlength}{\svgwidth}%
  \fi%
  \global\let\svgwidth\undefined%
  \global\let\svgscale\undefined%
  \makeatother%
  \begin{picture}(1,5.89808633)%
    \lineheight{1}%
    \setlength\tabcolsep{0pt}%
    \put(0.05,0){\includegraphics[width=\unitlength,page=1]{colorbar.pdf}}%
    \put(0.62,0.0){\makebox(0,0)[lt]{\lineheight{1.25}\smash{\begin{tabular}[t]{l}\footnotesize{0}\end{tabular}}}}%
    \put(0.62,2.75){\makebox(0,0)[lt]{\lineheight{1.25}\smash{\begin{tabular}[t]{l}\footnotesize{0.1}\end{tabular}}}}%
    \put(0.62,5.5){\makebox(0,0)[lt]{\lineheight{1.25}\smash{\begin{tabular}[t]{l}\footnotesize{0.22}\end{tabular}}}}%
    \put(-0.5,6.5){\makebox(0,0)[lt]{\lineheight{1.25}\smash{\begin{tabular}[t]{l}\small{$g_s$ \footnotesize [\si{\milli\m}]}\end{tabular}}}}%
  %  \put(0,0){\includegraphics[width=\unitlength,page=2]{colorbar_mufs.pdf}}%
  \end{picture}%
\endgroup%

		\end{subfigure}
		\caption{Evolution of the debonding of the BII: Value of the friction coefficient $\mu$ (top) and the sliding distance $\gs$ (bottom) on the contact area of the implant for different angles of rotation.}
		\label{img:friction_5deg}
	\end{figure}

\begin{figure}[H]
		\centering
		\begin{subfigure}[t]{0.47\textwidth}
			\input{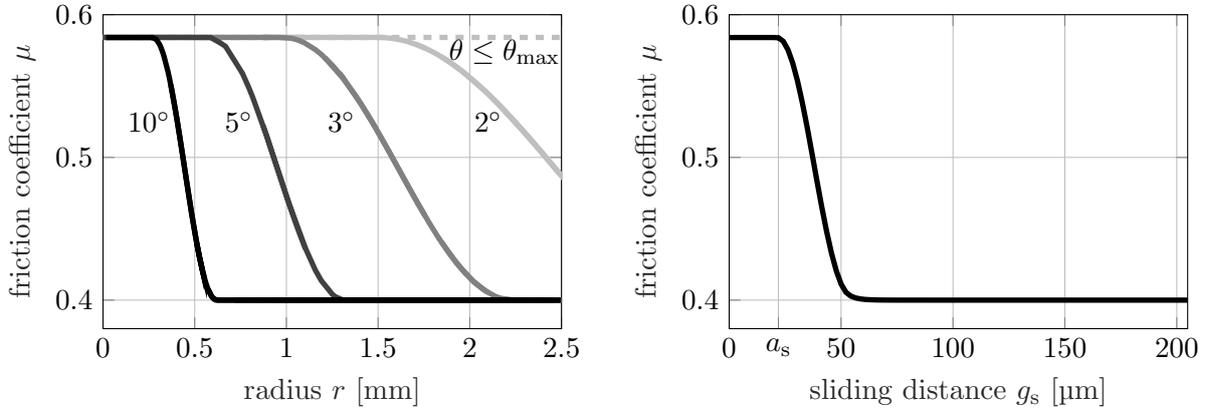}
			\caption{Variation of the friction coefficient $\mu$ as a function of the implant radius $r$ for different angles of rotation.}
			\label{img:mur}
		\end{subfigure}
		\hspace{5mm}
		\begin{subfigure}[t]{0.47\textwidth}
\begin{tikzpicture}

\begin{axis}[%
width=2.375in,
height=1.639in,
scale only axis,
xmin=0,
xmax=204.862250278451,
xlabel style={font=\color{white!15!black}},
xlabel={sliding distance $\gs$ [\si{\micro\meter}]},
xtick={0,22,50,100,150,200},
xticklabels={0,$\as$,50,100,150,200},
ymin=0.38,
ymax=0.6,
ylabel style={font=\color{white!15!black}},
ylabel={friction coefficient $\mu$},
ytick={0.4,0.5,0.6},
yticklabels={0.4,0.5,0.6},
axis background/.style={fill=white},
xmajorgrids,
ymajorgrids,
legend style={legend cell align=left, align=left, draw=white!15!black}
]
\addplot [color=black, line width=2.0pt]
  table[row sep=crcr]{%
0	0.584\\
0	0.584\\
0	0.584\\
1.16800091209291 	0.584\\
3.24698038736038 	0.584\\
5.33505735106079 	0.584\\
7.42459281442598 	0.584\\
9.51450883973138 	0.584\\
1.16046981127934e+01	0.584\\
1.36948868702128e+01	0.584\\
1.57850756138985e+01	0.584\\
1.78752640495568e+01	0.584\\
1.99654520805164e+01	0.584\\
2.2055707809626e+01	0.583805276752843\\
2.41469061454006e+01	0.5813481305375\\
2.6249011936119e+01	0.575496619061767\\
2.83677310829793e+01	0.566207146499539\\
3.05034451298262e+01	0.553754841489509\\
3.26549593986059e+01	0.53860929326889\\
3.48200808769698e+01	0.521400359502292\\
3.6995759803811e+01	0.502886675207112\\
3.91782931567286e+01	0.483913119382864\\
4.13635463445635e+01	0.465360152279986\\
4.35471872890082e+01	0.448088474175556\\
4.57249421721764e+01	0.432884107856662\\
4.78928860667826e+01	0.420410659428121\\
5.00476777803531e+01	0.411174367483197\\
5.21867410249023e+01	0.405504504502975\\
5.43083592553999e+01	0.402864575848439\\
5.64149688072341e+01	0.401580361713869\\
5.85145559730927e+01	0.400891726692256\\
6.06107697550416e+01	0.400490124648825\\
6.27052236297991e+01	0.400288147647757\\
6.4798406650778e+01	0.400169786311177\\
6.68909117598776e+01	0.400095913114151\\
6.89829405824734e+01	0.400052997029426\\
7.10745944771014e+01	0.400031370289012\\
7.31659757179469e+01	0.400016335896751\\
7.52572301129467e+01	0.400007933282899\\
7.73483562454626e+01	0.40000516225478\\
7.94393556527351e+01	0.400003360671109\\
8.15302739696813e+01	0.400001990650459\\
8.3621111487142e+01	0.40000099050139\\
8.57118757885199e+01	0.400000358701684\\
8.78025738659625e+01	0.400000100075368\\
8.98932108316263e+01	0.400000057841298\\
9.19838035851086e+01	0.400000027634179\\
9.40743712457619e+01	0.40000000884273\\
9.61649209679202e+01	0.400000001567985\\
9.82554502182245e+01	0.400000001065101\\
100.345959293898	0.40000000068643\\
102.436449858481	0.400000000388722\\
104.526922910318	0.400000000174352\\
106.617380132397	0.400000000044669\\
108.707820288797	0.40000000000003\\
110.798242095371	0.4\\
112.888649025209	0.4\\
114.979044764998	0.4\\
117.06942866663	0.4\\
119.159800401584	0.4\\
121.250162432665	0.4\\
123.340520264815	0.4\\
125.430873974261	0.4\\
127.521223168111	0.4\\
129.611567916864	0.4\\
131.701907780428	0.4\\
133.792243770168	0.4\\
135.882576317022	0.4\\
137.972905227677	0.4\\
140.063230134895	0.4\\
142.153550901148	0.4\\
144.243867218891	0.4\\
146.334179790271	0.4\\
148.424489109963	0.4\\
150.514795710049	0.4\\
152.605100713086	0.4\\
154.695403969939	0.4\\
156.78570573593	0.4\\
158.876005431211	0.4\\
160.966303167015	0.4\\
163.056598749772	0.4\\
165.146892195533	0.4\\
167.237183339192	0.4\\
169.327472117416	0.4\\
171.417759329525	0.4\\
173.508045408221	0.4\\
175.598330263788	0.4\\
177.688614388108	0.4\\
179.778898070289	0.4\\
181.869181240109	0.4\\
183.959463893958	0.4\\
186.049745943337	0.4\\
188.14002734254	0.4\\
190.23030806967	0.4\\
192.320588135102	0.4\\
194.410867318223	0.4\\
196.501145755362	0.4\\
198.591423333575	0.4\\
200.681699922127	0.4\\
202.771975587437	0.4\\
204.862250278451	0.4\\
};
\end{axis}

\end{tikzpicture}%
			\caption{Variation of the friction coefficient $\mu$ as a function of the total sliding distance $\gs$ at $r = R$.}
			\label{img:mu_gs}
		\end{subfigure}
		\caption{Evolution of the debonding of the BII: Behavior of the friction coefficient and the transition zone for the first data set.}
		\label{img:crack}
	\end{figure}

	The top row in \fig{img:friction_5deg} shows the distribution of the friction coefficient $\mu$ in the contact area for different angles of rotation. A transition zone (characterized by $\muk < \mu < \mus$), which may be understood as a crack front, cf. Ref. \cite{mathieu2012} and \fig{img:exp}, appears at \ang{1.13}. 
	The transition zone propagates inward in the radial direction from the external radius $R$ into the center, which corresponds to the crack mode assumed by the analytical models. 
	The transition zone can also be observed in \fig{img:mur}, which shows the value of the friction coefficient as a function of the implant radius for different angles of rotation.
	When the rotation angle increases, the width of the transition zone decreases. 
	It becomes apparent, that for the numerical model no full debonding is achieved after a rotation of \ang{10}, since for the nodes close to the center of the implant's contact surface the appropriate sliding distance to start the transitioning of $\mu$ has not been reached yet.
	In fact, under perfect twisting conditions, $\gs$ remains zero at the center of the implant.
	Thus the center of the implant will never debond for (perfect) twisting.
	
	The bottom row of \fig{img:friction_5deg} shows the distribution of the sliding distance over the contact surface for different angles of rotation. 
	Although the body starts sliding before a twisting angle of \ang{1}, the factor $\as$ prevents a change of the friction coefficient until a sliding distance of \SI{22}{\micro\m} is reached. 
	This is also shown in \fig{img:mu_gs}, which illustrates the variation of the friction coefficient as a function of the sliding distance $\gs$ for $r=R$ and data set 1. 
	The friction coefficient stays constant as $\mu = \mus$ for $\gs \le \as$ and then decreases, until it reaches $\mu = \muk$ at $\gs = \SI{66}{\micro\metre}$.	
		
	%---------------------------------------------------
	%---------------------------------------------------
	%--------------------------------------------------------------------
	%--------------------------------------------------------------------	
	%--------------------------------------------------------------------
	%--------------------------------------------------------------------
	%
	%--------------------------------------------------------------------
	% Work and energy
	%--------------------------------------------------------------------
	\subsection{Work of adhesion} \label{s:res_work}
	Due to the poor agreement between the experimental results and the analytical model developed in Mathieu~\etal~\cite{mathieu2012}, an energetic approach was proposed to determine the dissipated frictional energy $\Wfric$, the work of adhesion $\Wadh$, and the total energy necessary for debonding $\Wdeb$,\footnote{The strain energy inside the bodies, which generally should also be accounted for, is negligible in this case.} which are given by
	\begin{equation}
		\begin{aligned}
			\Wdeb &= \int_{\theta = \ang{0}}^{\theta = \ang{10}} T(\theta) \dd\theta, \quad
			\Wfric &= \int_{\theta = \thetamax}^{\theta = \ang{10}} \Tinf \dd\theta, \quad
			\Wadh &= \Wdeb - \Wfric.
		\end{aligned}
	\end{equation}
	Based on the experimental results, $\theta = \ang{10}$ was chosen to be the angle of rotation where the implant was completely debonded from the bone, indicated by a constant torque $\Tinf$. 
	The area-specific average work of adhesion $\Eadh$ is then given by
	\begin{equation} \label{eq:wadh}
	 \Eadh = \frac{\Wadh}{\pi R^2 \phiog},
	\end{equation}
	where $\phiog$ is the average of the initially osseointegrated area, see also \sect{s:res_osseo}. A visual analysis of the implant surfaces after debonding yielded $\phiog = 0.73$ for the first data set (see \fig{img:pattern1}) and $\phiog = 0.72$ for the second.
	
	\tab{tab:work} gives the results for the different energies with respect to the values of $\phiog$ determined by the surface analysis. 
	In the cases presented here, the analytical and the numerical models use $\phio(\bx)=1\,\forall\,\bx$. Therefore, the true area-specific work of adhesion for these models $\Eadh^*$ was computed by using $\phiog = 1$ in Eq.~\eqref{eq:wadh}. Results for $\Eadh$ where $\phiog\neq 1$ is used during the simulation are presented in \sect{s:res_osseo}. 
	
	For both data sets, the analytical solution by Mathieu~\etal~underestimates the total debonding work and the area-specific adhesion work, while the numerical results with the modified Coulomb's law lead to very good agreement with the experimental data. 
	The analytical solution with the modified Coulomb's law yields less accurate results for the second data set.
	
	For simplicity, the results in the remaining part of this work were obtained based only on the estimated parameters for the first data set and $\muk = 0.4$ (see \tab{tab:param_estim2}).
	%--------------------------------------------------------------------
	\begin{table}[H]
		\centering
		\begin{tabular}{|c|c||c|c|c|c|c|}
			\hline 
			data & \multirow{2}{*}{model} & \multirow{2}{*}{$\Wdeb$ [\si{\N\m}]} & \multirow{2}{*}{$\Wfric$ [\si{\N\m}]} & \multirow{2}{*}{$\Wadh$ [\si{\N\m}]} & \multirow{2}{*}{$\Eadh$ [\si{\N\per\m}]} & \multirow{2}{*}{$\Eadh^*$ [\si{\N\per\m}]}\\ 
			set & & & & & & \\
			\hline \hline
			\multirow{4}{*}{1} 
			& \multicolumn{1}{c||}{exp.} & 0.0070 & 0.0056 & 0.0014 & 98 & 71 \\
			& \multicolumn{1}{c||}{ana.~Ref.~\cite{mathieu2012}} & 0.0066 & 0.0056 & 0.0010 & 70 & 51 \\
			& \multicolumn{1}{c||}{ana.~Eq.~\eqref{eq:new_ana}} & 0.0071 & 0.0057 & 0.0014 & 98 & 72 \\ 
			& \multicolumn{1}{c||}{sim.} & 0.0071 & 0.0057 & 0.0014 & 98 & 72 \\ 
			\hline
			\multirow{4}{*}{2} 
			& \multicolumn{1}{c||}{exp.} & 0.0088 & 0.0060 & 0.0028 & 198 & 143\\ 
			& \multicolumn{1}{c||}{ana.~Ref.~\cite{mathieu2012}} & 0.0080 & 0.0064 & 0.0016 & 120 & 81\\ 
			& \multicolumn{1}{c||}{ana.~Eq.~\eqref{eq:new_ana}} & 0.0087 & 0.0067 & 0.0020 & 141 & 102\\ 
			& \multicolumn{1}{c||}{sim.} & 0.0089 & 0.0062 & 0.0027 & 191 & 138\\ 
			\hline
		\end{tabular}
		\caption{Total debonding energy $\Wdeb$, frictional energy $\Wfric$, work of adhesion $\Wadh$, and corresponding area-specific works of adhesion $\Eadh$ and $\Eadh^*$ for the different models and data sets.}
		\label{tab:work}
	\end{table}
	%--------------------------------------------------------------------
	%
	%--------------------------------------------------------------------
	% Osseointegration
	%--------------------------------------------------------------------
	\subsection{Partial osseointegration} \label{s:res_osseo}		
\begin{figure}[htbp]
		\centering
		\begin{subfigure}[t]{0.24\textwidth}
			%	\centering
			\includegraphics[width=1.1\textwidth]{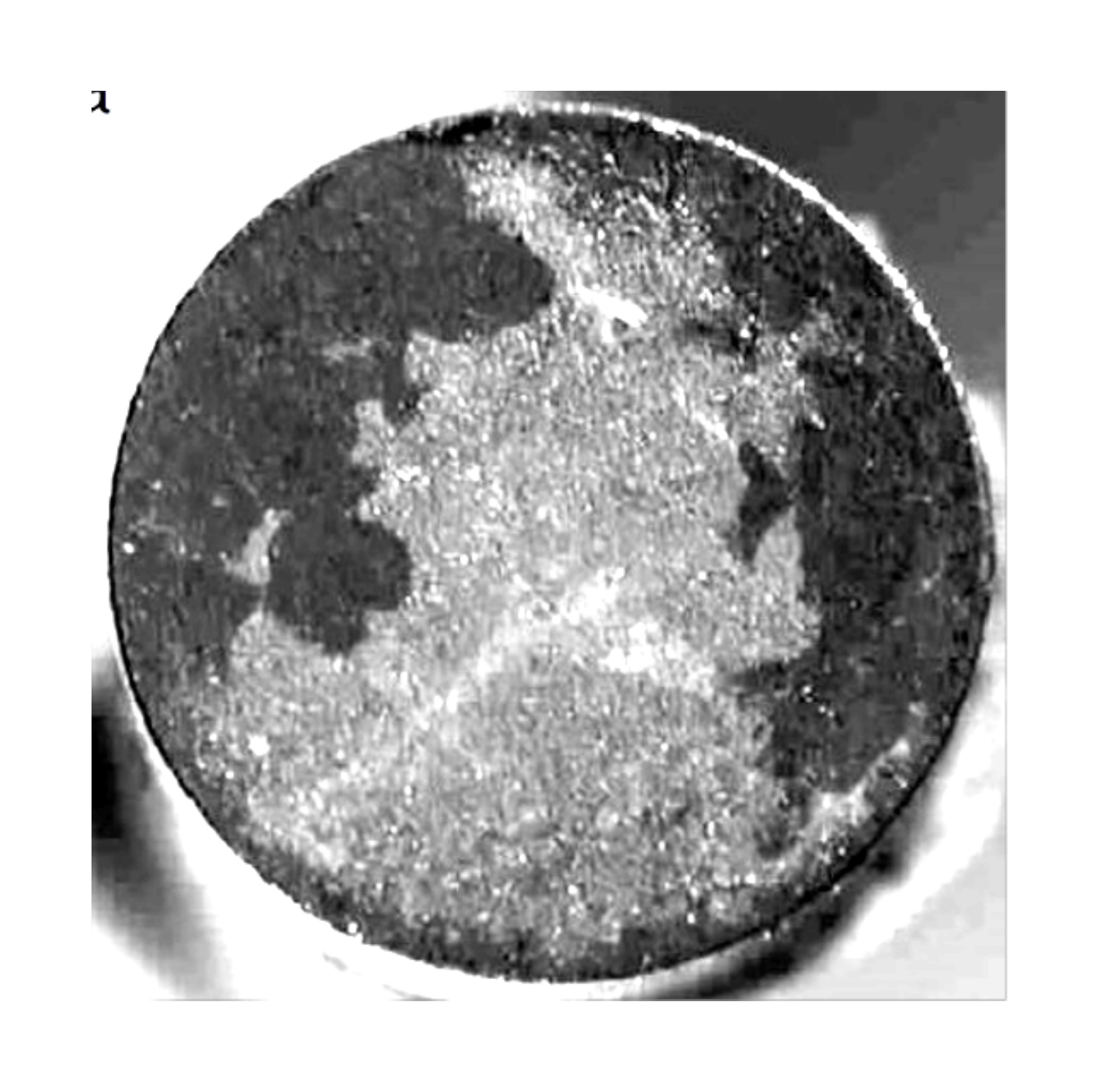}
			\caption{Original pattern \cite{mathieu2012}.}
			\label{img:pattern1}
		\end{subfigure}
		\begin{subfigure}[t]{0.24\textwidth}
			\centering
			\includegraphics[width=0.95\textwidth]{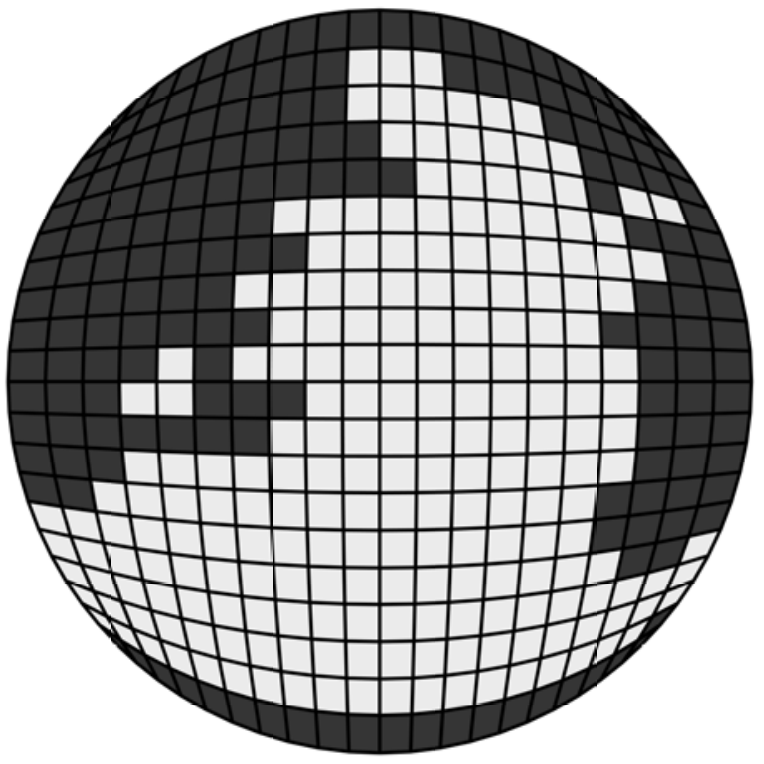} 
			\caption{Reconstructed pattern.}
			\label{img:pattern2}
		\end{subfigure}
		\begin{subfigure}[t]{0.24\textwidth}
			\centering
			\includegraphics[width=0.95\textwidth]{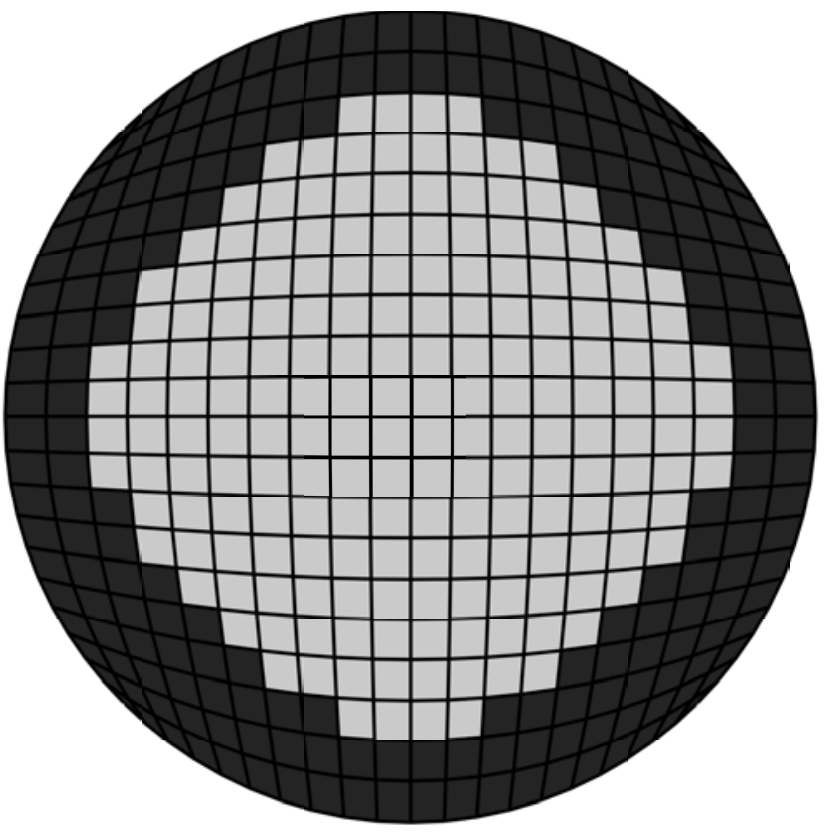}
			\caption{Circular pattern.}
			\label{img:pattern3}
		\end{subfigure}	
		\begin{subfigure}[t]{0.24\textwidth}
			\centering
			\includegraphics[width=0.95\textwidth]{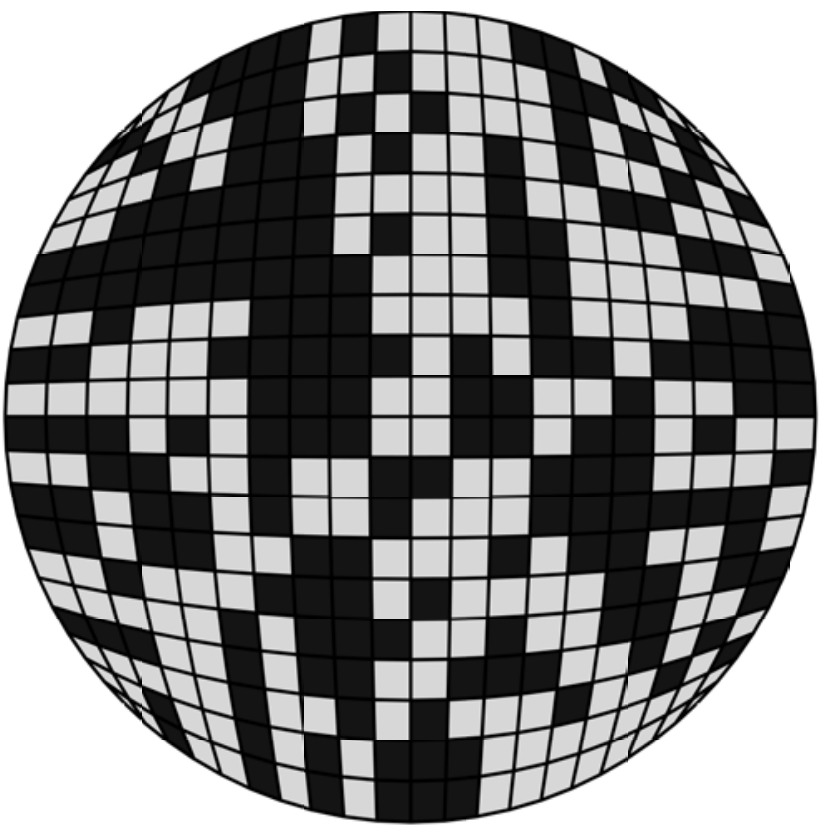}
			\caption{Random pattern.}
			\label{img:pattern4}
		\end{subfigure}
		\caption{Original and constructed osseointegration patterns with average osseointegration $\phiog=0.55$. Light areas represent full osseointegration ($\phio=1$), while dark areas represent no osseointegration ($\phio=0$). For the corresponding elements in the reconstructed and artificial patterns $\phio^e$ is set to 1.}
		\label{img:osseo_patterns}
	\end{figure}

	Mathieu~\etal~\cite{mathieu2012} showed that part of the limitations of their model lies in the assumption of a fully bonded surface at the beginning of the experiment. 
	The analysis of the implants' surfaces after debonding indicated that most likely, full osseointegration was not achieved. 
	This resulted in regions where no bone tissue was initially attached to the implant surface, as seen in \fig{img:pattern1}, and thus no adhesive or frictional effects can take place. 
	This state of partial osseointegration can be readily investigated with the proposed numerical model. Therefore, an analysis assuming inhomogeneous initial bonding was performed, with various distributions $\phio(\bx)$ in model \eqref{eq:phi}.

	In order to analyze the debonding behavior of the BII and to determine the influence of the percentage and the distribution of osseointegration, different cases for a fine mesh with 400 contact elements on the implant surface were constructed: 
	First, a bonding pattern based on \fig{img:pattern1}, assuming that light gray areas indicate osseointegration, was reconstructed. 
	From a visual inspection of the photograph, the sample in \fig{img:pattern1} is considered to have average osseointegration $\phiog = 0.55$.%\footnote{In Ref.~\cite{mathieu2012}, they considered $\phiog=0.73$.} 
	Second, to compare the influence of osseointegration patterns, two other patterns with $\phiog=0.55$ were constructed. 
	A simple circular pattern, where only the center part of the interface is integrated and last, a random distribution. 
	For patterns (b)--(d), $\phio=1$ within the light gray surface elements. The corresponding osseointegration patterns are shown in \fig{img:osseo_patterns}.
	%--------------------------------------------------------------------
	%--------------------------------------------------------------------
		%
		%--------------------------------------------------------------------
		% Partial osseointegration for for different material parameters
		%--------------------------------------------------------------------
		\subsubsection{Partial osseointegration for different material parameters}
\begin{figure}[H]
			\centering
			\begin{overpic}[width=0.175\linewidth]{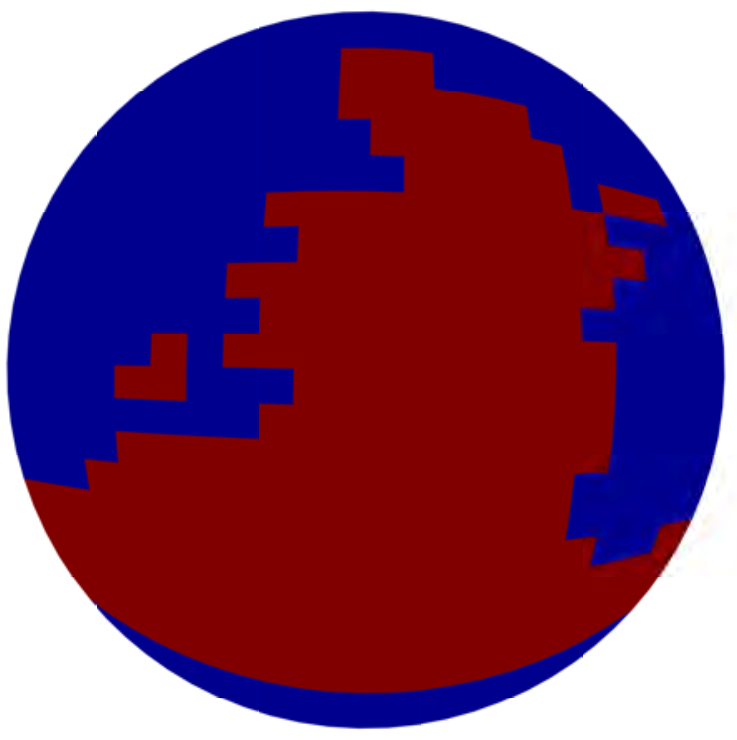}
				\put (90,5) {\large$\ang{1}$}
			\end{overpic}
			\begin{overpic}[width=0.175\linewidth]{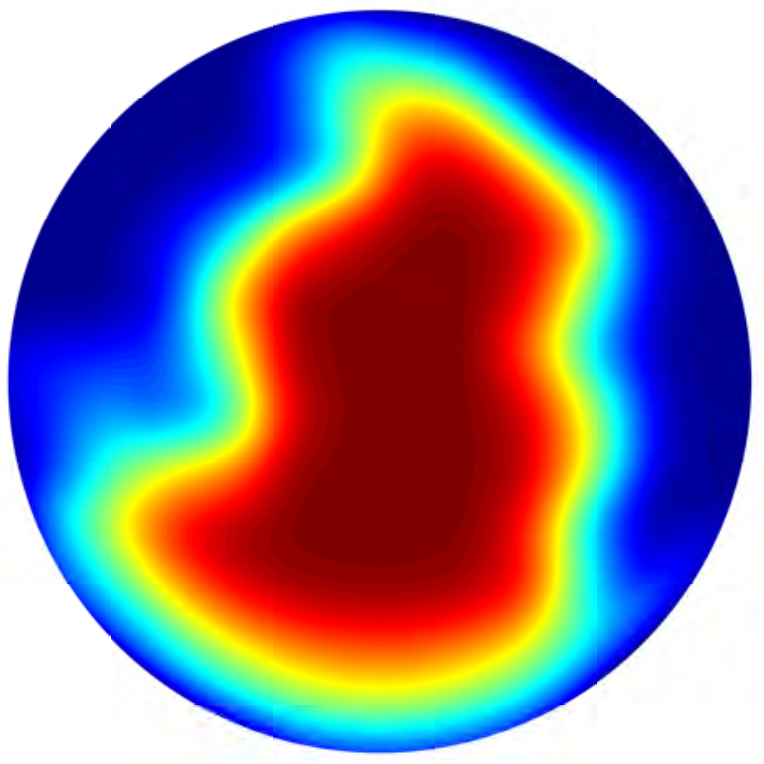}
				\put (90,5) {\large$\ang{2}$}
			\end{overpic}
			\begin{overpic}[width=0.175\linewidth]{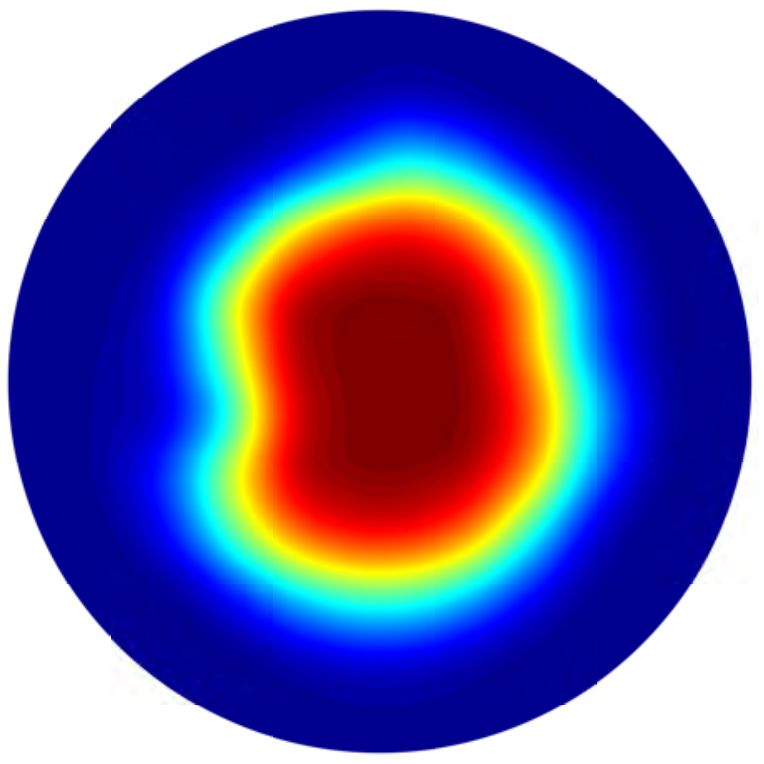}
				\put (90,5) {\large$\ang{3}$}
			\end{overpic}
			\begin{overpic}[width=0.175\linewidth]{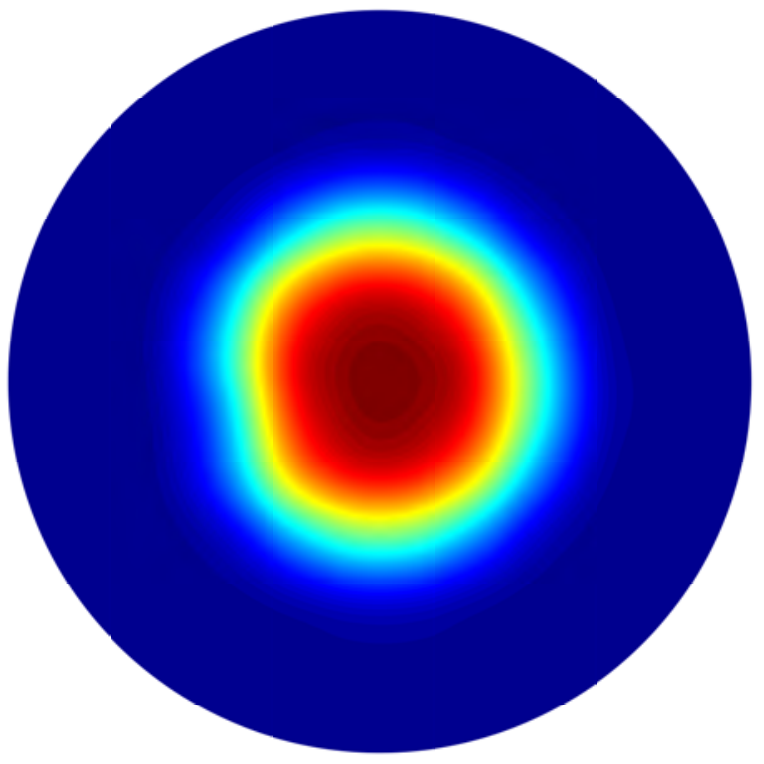}
				\put (90,5) {\large$\ang{4}$}
			\end{overpic}
			\begin{overpic}[width=0.175\linewidth]{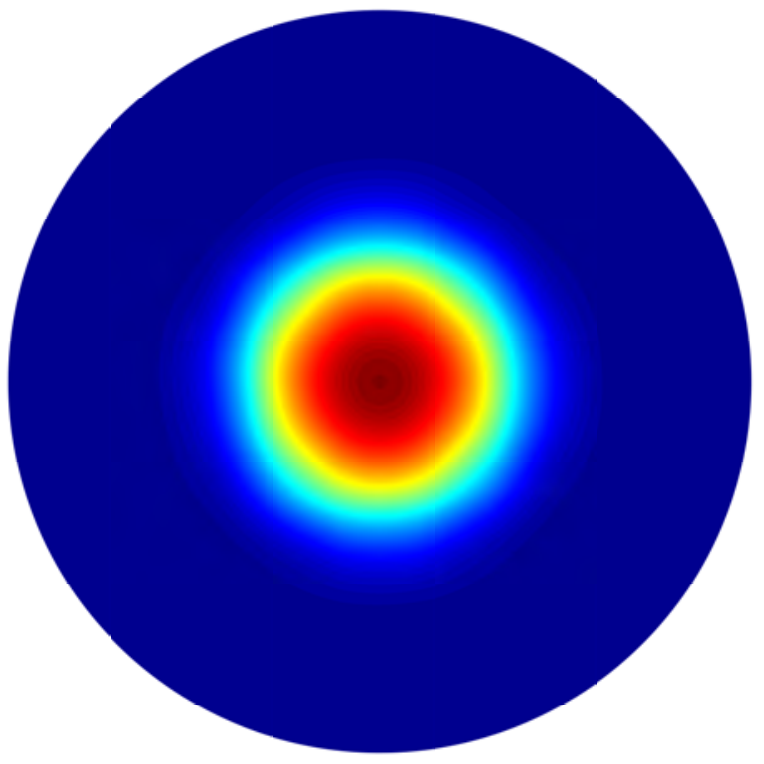}
				\put (90,5) {\large$\ang{5}$}
			\end{overpic}
			\begin{subfigure}{0.05\linewidth}
				\vspace{-25mm}
				\def\svgwidth{0.45\textwidth}
%% Accompanies image file 'colorbar_mufs.pdf' (pdf, eps, ps)
\begingroup%
  \makeatletter%
  \providecommand\color[2][]{%
    \errmessage{(Inkscape) Color is used for the text in Inkscape, but the package 'color.sty' is not loaded}%
    \renewcommand\color[2][]{}%
  }%
  \providecommand\transparent[1]{%
    \errmessage{(Inkscape) Transparency is used (non-zero) for the text in Inkscape, but the package 'transparent.sty' is not loaded}%
    \renewcommand\transparent[1]{}%
  }%
  \providecommand\rotatebox[2]{#2}%
  \newcommand*\fsize{\dimexpr\f@size pt\relax}%
  \newcommand*\lineheight[1]{\fontsize{\fsize}{#1\fsize}\selectfont}%
  \ifx\svgwidth\undefined%
    \setlength{\unitlength}{44.31712735bp}%
    \ifx\svgscale\undefined%
      \relax%
    \else%
      \setlength{\unitlength}{\unitlength * \real{\svgscale}}%
    \fi%
  \else%
    \setlength{\unitlength}{\svgwidth}%
  \fi%
  \global\let\svgwidth\undefined%
  \global\let\svgscale\undefined%
  \makeatother%
  \begin{picture}(1,5.89808633)%
    \lineheight{1}%
    \setlength\tabcolsep{0pt}%
    \put(0,0){\includegraphics[width=\unitlength,page=1]{colorbar.pdf}}%
    \put(0.0,6.5){\makebox(0,0)[lt]{\lineheight{1.25}\smash{\begin{tabular}[t]{l} \small $\mu$\end{tabular}}}}%
    \put(0.54832158,5.68163756){\makebox(0,0)[lt]{\lineheight{1.25}\smash{\begin{tabular}[t]{l}\small 0.8\end{tabular}}}}%
    \put(0.54832158,2.9){\makebox(0,0)[lt]{\lineheight{1.25}\smash{\begin{tabular}[t]{l}\small 0.6\end{tabular}}}}%
    \put(0.54832158,0.00307399){\makebox(0,0)[lt]{\lineheight{1.25}\smash{\begin{tabular}[t]{l}\small0.4\end{tabular}}}}%

    \put(0,0){\includegraphics[width=\unitlength,page=2]{colorbar.pdf}}%
  \end{picture}%
\endgroup%

			\end{subfigure}
			\begin{overpic}[width=0.175\linewidth]{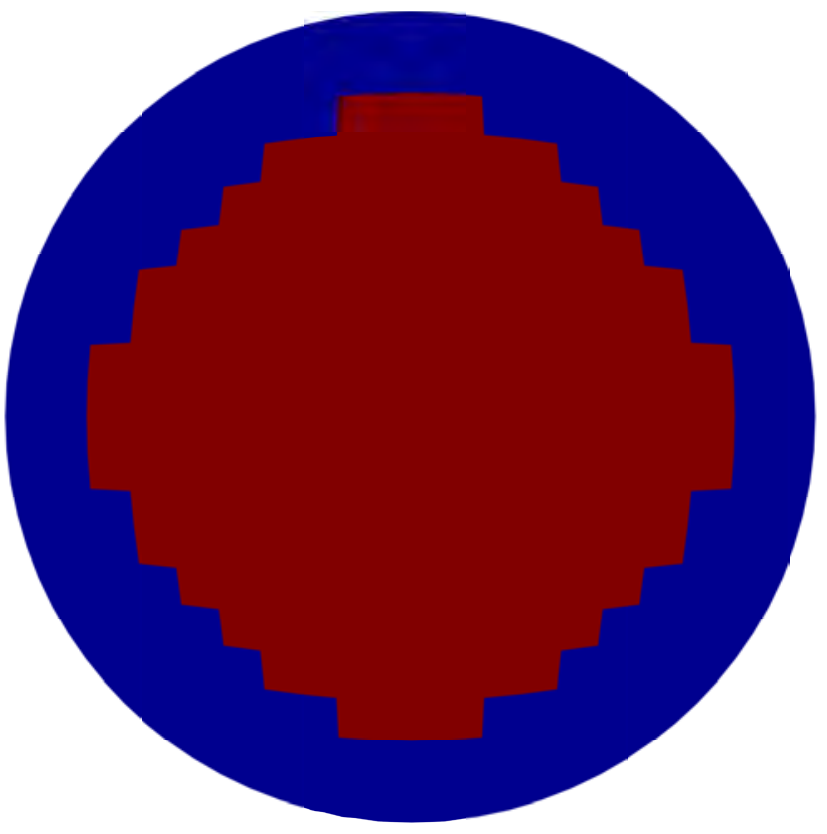}
				\put (90,5) {\large$\ang{1}$}
			\end{overpic}
			\begin{overpic}[width=0.175\linewidth]{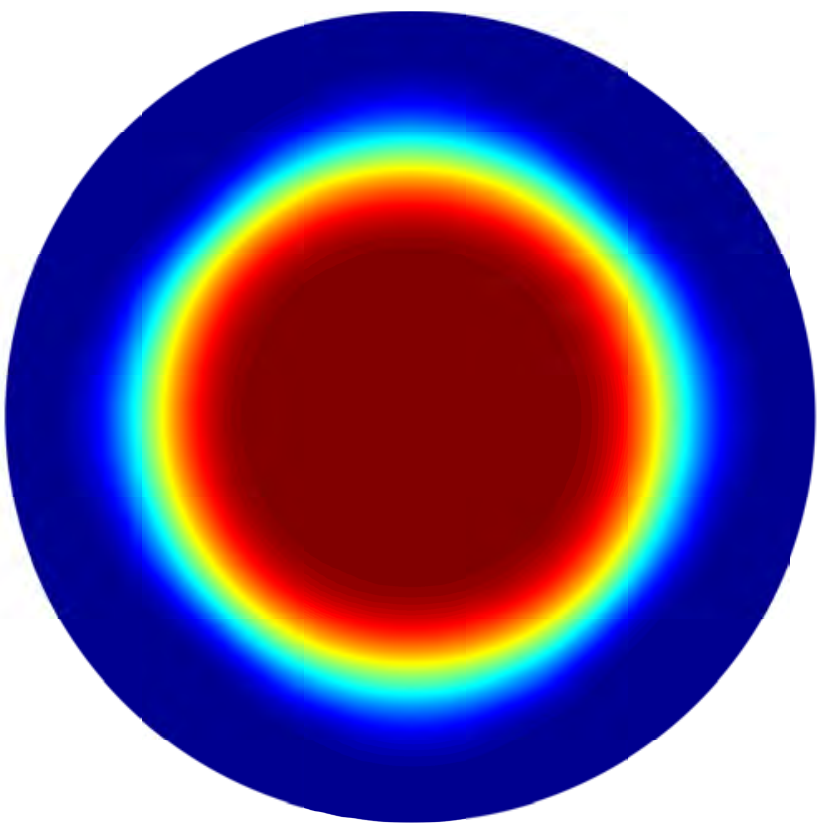}
				\put (90,5) {\large$\ang{2}$}
			\end{overpic}
			\begin{overpic}[width=0.175\linewidth]{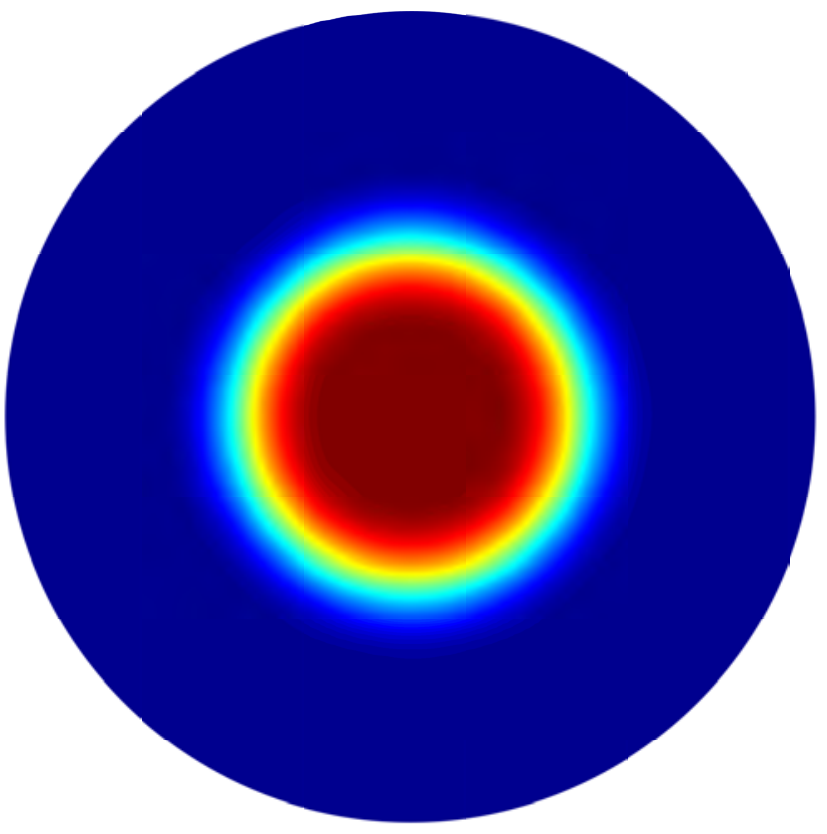}
				\put (90,5) {\large$\ang{3}$}
			\end{overpic}
			\begin{overpic}[width=0.175\linewidth]{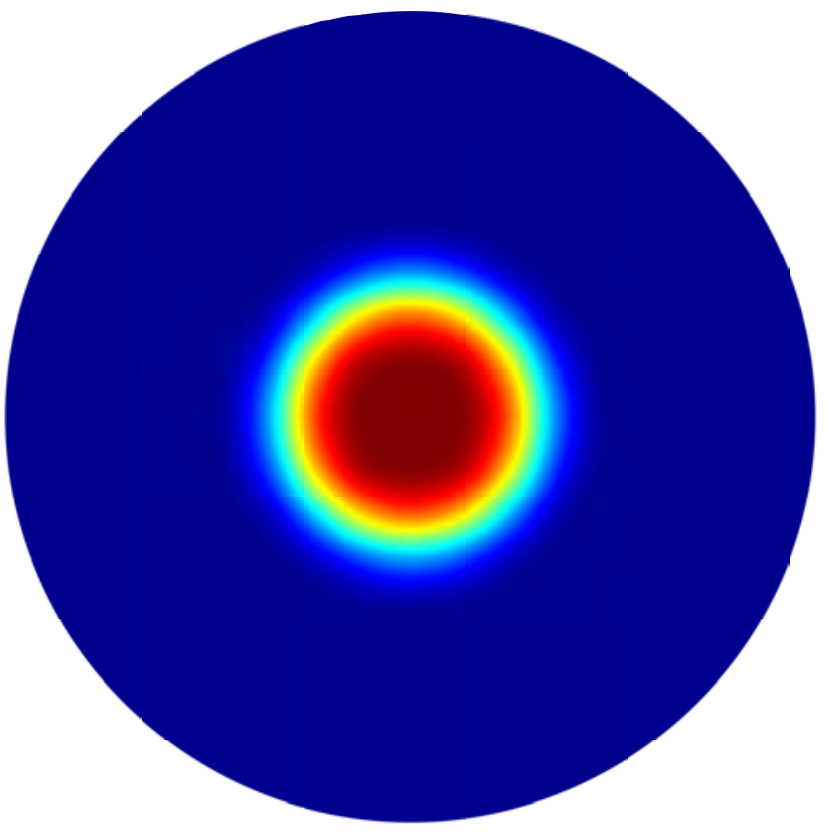}
				\put (90,5) {\large$\ang{4}$}
			\end{overpic}
			\begin{overpic}[width=0.175\linewidth]{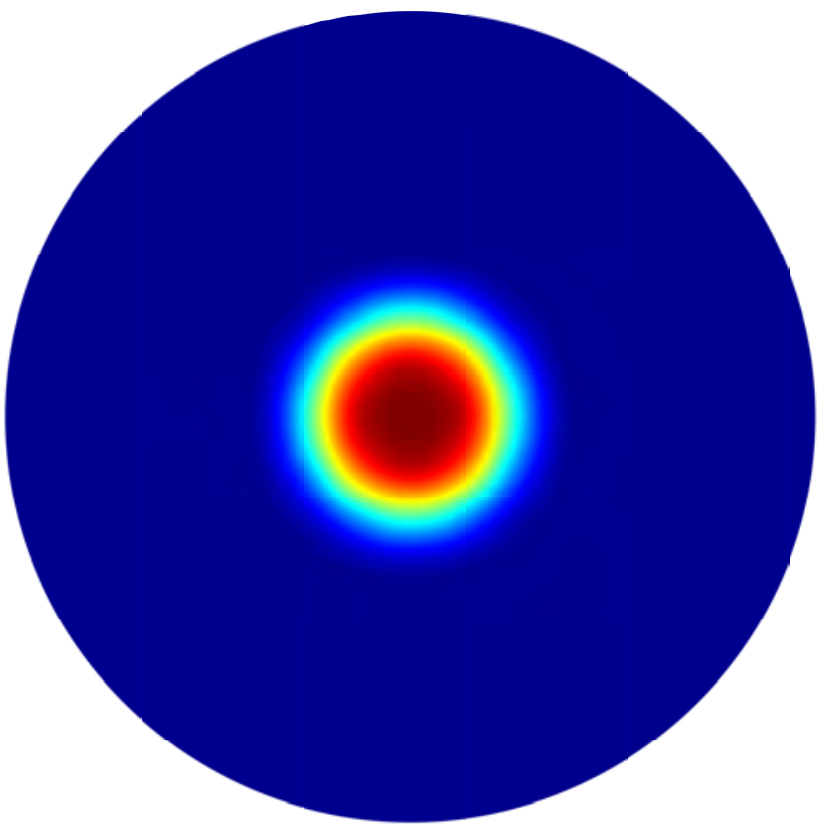}
				\put (90,5) {\large$\ang{5}$}
			\end{overpic}
			\begin{subfigure}{0.05\linewidth}
				\vspace{-25mm}
				\def\svgwidth{0.45\textwidth}
%% Accompanies image file 'colorbar_mufs.pdf' (pdf, eps, ps)
\begingroup%
  \makeatletter%
  \providecommand\color[2][]{%
    \errmessage{(Inkscape) Color is used for the text in Inkscape, but the package 'color.sty' is not loaded}%
    \renewcommand\color[2][]{}%
  }%
  \providecommand\transparent[1]{%
    \errmessage{(Inkscape) Transparency is used (non-zero) for the text in Inkscape, but the package 'transparent.sty' is not loaded}%
    \renewcommand\transparent[1]{}%
  }%
  \providecommand\rotatebox[2]{#2}%
  \newcommand*\fsize{\dimexpr\f@size pt\relax}%
  \newcommand*\lineheight[1]{\fontsize{\fsize}{#1\fsize}\selectfont}%
  \ifx\svgwidth\undefined%
    \setlength{\unitlength}{44.31712735bp}%
    \ifx\svgscale\undefined%
      \relax%
    \else%
      \setlength{\unitlength}{\unitlength * \real{\svgscale}}%
    \fi%
  \else%
    \setlength{\unitlength}{\svgwidth}%
  \fi%
  \global\let\svgwidth\undefined%
  \global\let\svgscale\undefined%
  \makeatother%
  \begin{picture}(1,5.89808633)%
    \lineheight{1}%
    \setlength\tabcolsep{0pt}%
    \put(0,0){\includegraphics[width=\unitlength,page=1]{colorbar.pdf}}%
    \put(0.0,6.5){\makebox(0,0)[lt]{\lineheight{1.25}\smash{\begin{tabular}[t]{l}$\mu$\end{tabular}}}}%
    \put(0.54832158,5.68163756){\makebox(0,0)[lt]{\lineheight{1.25}\smash{\begin{tabular}[t]{l}\small 0.97\end{tabular}}}}%
    \put(0.54832158,2.9){\makebox(0,0)[lt]{\lineheight{1.25}\smash{\begin{tabular}[t]{l}\small 0.69\end{tabular}}}}%
    \put(0.54832158,0.00307399){\makebox(0,0)[lt]{\lineheight{1.25}\smash{\begin{tabular}[t]{l}\small 0.4\end{tabular}}}}%
  \end{picture}%
\endgroup%

			\end{subfigure}
			\caption{Variation of the friction coefficient $\mu$ at the BII for different angles of rotation for different patterns of partial osseointegration.} 
			\label{img:osseo_5deg}
		\end{figure}

\begin{figure}[H]	
			\centering
			\input{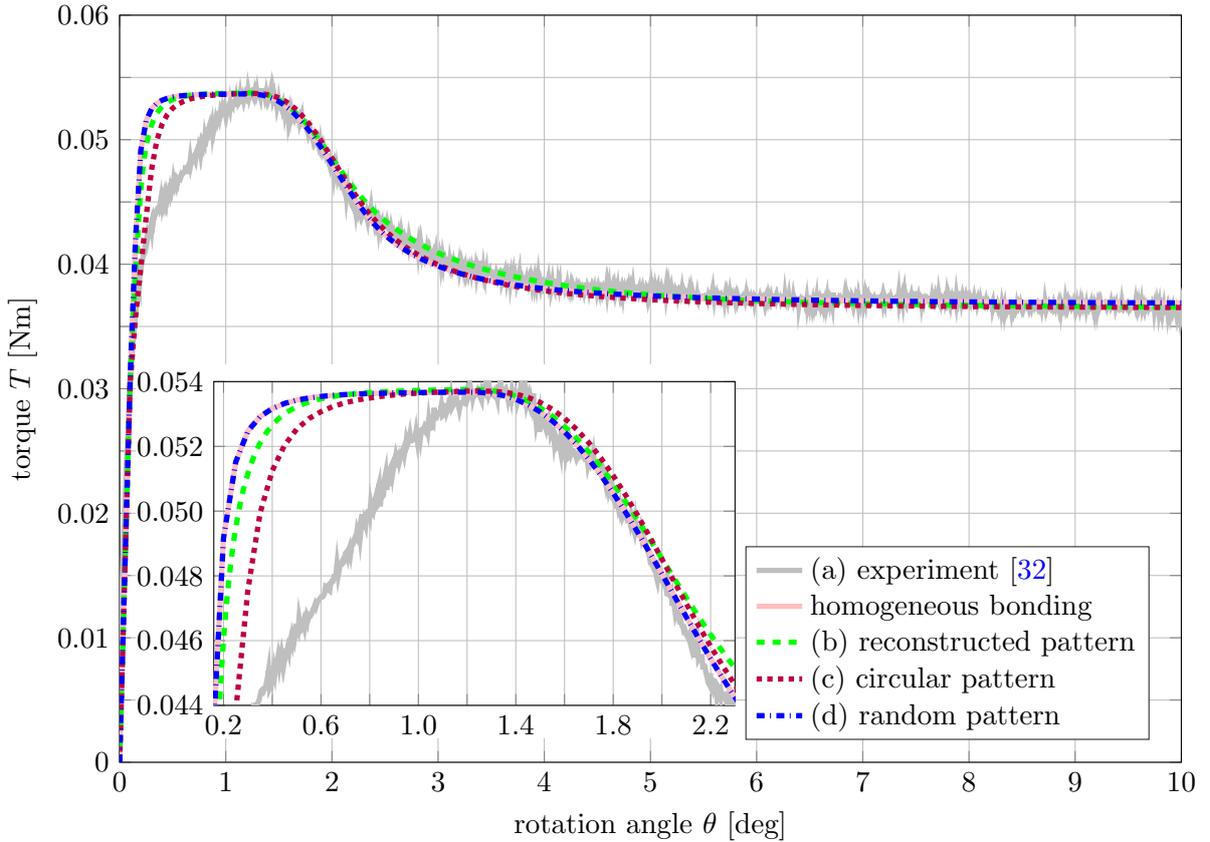}
			\caption{Partial osseointegration: Variation of the Torque $T$ as a function of the imposed rotation angle $\theta$ for different cases of the initial bonding state of the interface with the material parameters shown in \tab{tab:osseo_param}.}
			\label{img:torque_a_inhom}
		\end{figure}

		When the model parameters are fitted for every pattern (see \tab{tab:osseo_param}), the torque-angle curve only weakly depends on the various distributions of osseointegration, as seen in \fig{img:torque_a_inhom}, resulting in a minimum error of \SI{1.95}{\percent} for the reconstructed pattern, \SI{1.99}{\percent} for the circular pattern and \SI{2.15}{\percent} for the random distribution, respectively. 
		While the error is slightly larger compared to the reconstructed pattern, the circular pattern has a closer fit to the experimental data for $\theta <\thetamax $, which leads to a small improvement of the prediction of $\Eadh$. 
		Since the model parameters are unknown, fitting the parameters to the experimental data leads to similar curves for every presented case. 
		It can be seen, that out of the different osseointegration patterns, the random pattern is the closest to the results obtained with full initial bonding. 
		The random pattern still has a balanced degree of osseointegration over the implant radius, while pattern (b) and (c) are only weakly (or not) bonded on the outer part of the implant and thus, a bigger difference is seen in the beginning of the torque curve. 
		Overall, the aforementioned results lead to the conclusion that without knowing the friction coefficients, the torque-per-angle curve does not provide sufficient information on the degree and distribution of osseointegration of the BII.
		%--------------------------------------------------------------------
		\begin{table}[H]
			\centering
			\begin{tabular}{|c||c|c|c|c|c|c|c|c|}
				\hline 
				bonding & \multirow{2}{*}{$\phiog$} &\multirow{2}{*}{$d$ [\si{\micro\m}]}  & \multirow{2}{*}{$\mus$} & \multirow{2}{*}{$\as$ [\si{\micro \m}]}  & \multirow{2}{*}{$\bs$}  & \multirow{2}{*}{$\errmp$ [\si{\percent}]} & \multirow{2}{*}{$\Wadh$ [\si{\N\m}]} & \multirow{2}{*}{$\Eadh$ [\si{\N\per\m}]}\\ 
				pattern & & & & & & & & \\
				\hline \hline 
				homog. & 1 & 4.90 & 0.58 & 22 & 0.74 & 2.240 &  0.0014 & 130 \\
				(b)    & 0.55 & 4.80 & 0.80 & 21 & 0.63 & 1.949 &  0.0015 & 139 \\
				(c)	   & 0.55 & 4.80 & 0.97 & 15 & 0.67 & 1.988 &  0.0014 & 130 \\
				(d)    & 0.55 & 4.85 & 0.73 & 22 & 0.66 & 2.152 &  0.0015 & 139 \\
				\hline
			\end{tabular} 
			\caption{Change in model parameters and results for implants with partial initial bonding compared to homogeneous bonding (see \fig{img:osseo_patterns}).}
			\label{tab:osseo_param}
		\end{table}	
		%--------------------------------------------------------------------
		As the state of an element does not depend on the states of neighboring elements, the total sliding distance is not affected by inhomogeneous initial bonding. 
		Therefore, a similar propagation of the transition zone as in the homogeneous case can be seen in \fig{img:osseo_5deg}.
		In contrast to the torque, partial osseointegration has a larger effect on the model parameters, as shown in \tab{tab:osseo_param}. In general, partial osseointegration leads to the estimation of higher $\mus$, that are still well within the range of reported values in the literature. 
		In addition, the transition time is notably lower and for the distinct patterns (b) and (c), the sliding threshold is lower, as well.
		%--------------------------------------------------
		%--------------------------------------------------------------------
		%--------------------------------------------------------------------
		%--------------------------------------------------------------------
		%
		%--------------------------------------------------------------------
		% Partial osseointegration for equal material parameters
		%--------------------------------------------------------------------
		\subsubsection{Partial osseointegration for equal material parameters}
\begin{figure}[H]	
			\centering
			\input{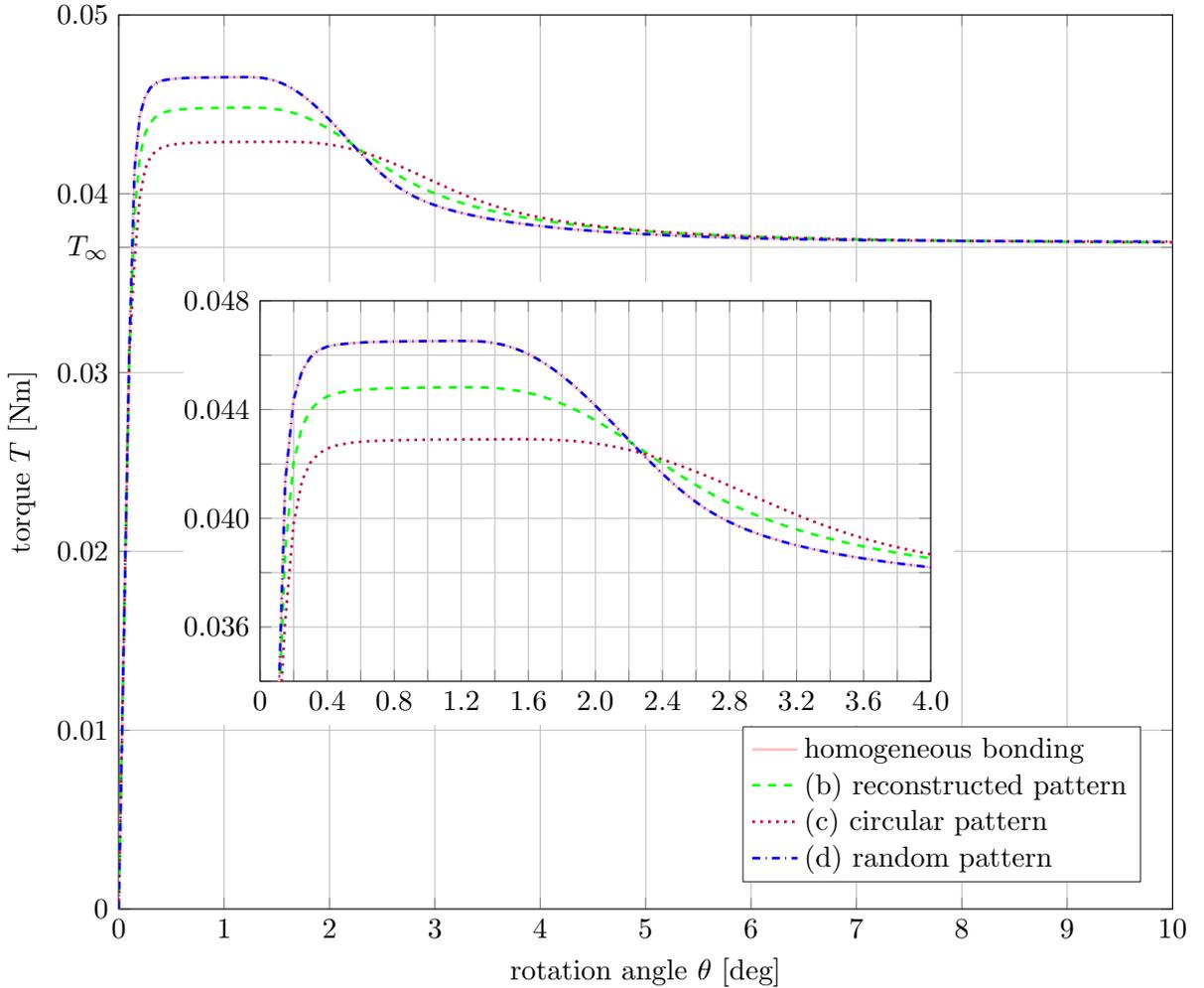}
			\caption{Partial osseointegration: Variation of the Torque $T$ as a function of the imposed rotation angle $\theta$ for different patterns of the initial bonding state of the interface with $\phiog=0.55$ and the same material parameters for all patterns. $\Tinf=\SI{0.0368}{\N\m}$ according to data set 1 (see \tab{tab:datasets}).}
			\label{img:torque_a_inhom_noscale}
		\end{figure}

		The effect of the different osseointegration patterns becomes much more distinct when studying the results obtained for equal material parameters for bone tissue and the implant, as presented in \fig{img:torque_a_inhom_noscale}. Here, the parameters for the homogeneous case of the first data set ($\muk=0.4$) were used (see \tabs{tab:param_estim1} and \ref{tab:param_estim2}) with $\phiog=0.55$ for all four patterns. Using partial osseointegration only affects the part of the deformation where (tangential) adhesive forces are expected to play a mayor role, represented by the peak in the torque curve. In general, partial osseointegration patterns lead to a lower $\Tmax$ and differences in the softening of the curve. The initial slope of the torque curve and $\Tinf$ are not affected by the different osseointegration patterns. Applying $\phio=0.55$ to all contact elements in the homogeneous bonding case produces the same result as using a randomly distributed pattern with $\phiog=0.55$, where $\phio=1$ for the osseointegrated elements. The distinct patterns (b) and (c) with weak or no boding on the outer part of the implant produce a lower peak and longer softening periods.
		%--------------------------------------------------------------------
		%--------------------------------------------------------------------
	%
	%--------------------------------------------------------------------
	% Discussion
	%--------------------------------------------------------------------
	\subsection{Discussion and Limitations} \label{s:res_disc}
	This work aims to provide a circular friction law to model the debonding of the BII. The mechanical model is incorporated into an analytical model and is implemented into an accurate and efficient contact algorithm with NURBS-enriched finite elements and thus, allows to predict the debonding of osseointegrated implants numerically. The model only depends on four physiological parameters ($\mus,\,p,\,\as,$ and $\bs$), that can be determined by a parameter study based on experimental results. Moreover, it allows for reasonable estimations of other parameters, such as the bone's Young's modulus and adhesion energy. However, due to the limited experimental data analyzed here, no \textit{a-priori} determination of the input parameters can be made yet. While the analytical solution of the proposed friction law already leads to good estimates, the results obtained by the finite element simulation are more accurate as they can also account for non-linear material behavior, large deformations, and partial osseointegration.
	
	One difficulty associated with the present study arises from the contact pressure, which is required to establish frictional contact. In Mathieu~\etal~\cite{mathieu2012}, it was reported that although the experimental pressure was minimized, it could not be completely excluded nor measured. Since the experimental torque does not go to zero for fully debonded implants, a normal pressure is likely to remain applied to the implants. As friction coefficients and normal force are unknown, no statement can be made about the accuracy of the estimated contact pressure. Furthermore, in the beginning, $p$ accounts for applied normal forces as well as adhesive forces due to chemical and mechanical bonding. The measurement or elimination of applied normal forces would clearly determine the friction coefficient for the broken state and thus also for the unbroken state of a certain sample. Therefore, an improvement of the testing machine used for the experimental measurements is needed.
	
	While the estimated parameters are within a reasonable range compared to the literature (see e.g. Refs.~\cite{damm2015,novitskaya2011,shirazi-adl1993}), and result in a good qualitative and quantitative representation of the torque-angle curve, there are still visible differences, in particular at the beginning of the peak. Following the argumentation of the reference study, these differences might be explained by the assumption of a full initial bonding between bone and implant, while an initial bonding of 30--70~\si{\percent} is reported in the literature~\cite{branemark1997,marin2010}. Accounting for inhomogeneous osseointegration in our model has shown an influence on the torque curves and the model parameters, such as the friction coefficient for the unbroken state. As these values were not or cannot be measured experimentally yet, it is assumed that the parameters obtained by incorporating imperfect osseointegration are more accurate. Furthermore, a partial bonding will most likely lead to a more complex crack front and propagation than what is assumed by the analytical and numerical models. 
	
	Other factors that were not taken into account in the present work are the roughness of bone and implant surfaces, as well as wear and debris. Furthermore, only a healing time of 7 weeks was considered, while a comparison of different healing times in terms material parameters and of osseointegration would be interesting. Thus, a study on the influence of the surface roughness and healing time is planned for future work.
	
	Another interesting aspect for future work is the application of the model to actual implant and bone geometries, for example in artificial hip joints and dental implants. The latter have been recently examined by Rittel~\etal~\cite{Rittel2018} to study the influence of partial osseointegration on implant stability and cohesive failure.
	In addition, only torsional debonding was tested in this work while other loading conditions, such as push-in and pull-out of the implant are more commonly analyzed (see e.g. Refs.~\cite{berahmani2015,bishop2014,damm2015,wennerberg2014}). 
	
	Although, especially trabecular bone is highly non-linear and anisotropic, the choice of an isotropic, non-linear elastic Neo-Hookean material model here leads to reasonable results for the planar mode III debonding of titanium and cortical bone, due to the small deformation. It has to be investigated if this still holds for, e.g. pull-out tests and contact with trabecular bone. Especially for the modeling of cohesive failure of bone a fracture model would be needed.
%	
%--------------------------------------------------------------------
%  Conclusion
%-------------------------------------------------------------------- 	
\section{Conclusion} \label{s:concl}
This paper presents an extension of the classical Coulomb's law to a non-constant, state-dependent friction coefficient and its application to the debonding of homogeneous and partially osseointegrated implants. An analytical model and numerical solutions with the new friction model are compared to experimental and analytical results of the reference study~\cite{mathieu2012}. Overall it is shown, that assuming a smooth transition from an unbroken to a broken state, characterized by a decreasing friction coefficient during the debonding process leads to good agreement with experimental data with both, an analytical and a numerical approach. While the analytical model is simple, it is an efficient way to provide initial estimates for this kind of experiment. The numerical results on the other hand are more accurate and allow for more complex material behavior, stress distribution, and (partial) osseointegration. Both approaches enable the estimation of several parameters of the BII. The proposed friction model is expected to help in understanding the debonding phenomena at the BII. An extension to adhesive friction \cite{mergel2019} and rough surfaces, as well as hip implants, will be considered in future works.
%	
%--------------------------------------------------------------------
%  Acknowledgements
%--------------------------------------------------------------------
\section*{Acknowledgments}
This work has received funding from the European Research Council (ERC) under the European Union’s Horizon 2020 research and innovation program (grant agreement No 682001, project ERC Consolidator Grant 2015 BoneImplant). This work was supported by the Jülich Aachen Research Alliance Center for Simulation
and Data Science (JARA-CSD) School for Simulation and Data Science (SSD).

\section*{Conflict of interest}
The authors declare that they have no conflict of interest.
%
%--------------------------------------------------------------------
% Appendix
%--------------------------------------------------------------------
\section*{Appendix}\label{s:appndx}
\appendix
\section{Contact formulation} \label{s:appndx_contact}
In this section the basic contact mechanics and the numerical discretization are briefly summarized. For the formulation of sliding contact, one can use elastoplasticity theory in order to incorporate friction response in the tangential direction (see e.g. Refs.~\cite{laursen2013,sauer2015,wriggers2006}). In the context of this work, we use another approach that is based on the surface potential-based contact formulations of Duong and Sauer~\cite{duong2019} and Sauer and de Lorenzis~\cite{sauer2013}. The reader is referred to Corbett and Sauer~\cite{corbett2014,corbett2015} for the NURBS-enriched contact finite element discretization.
	%
	%--------------------------------------------------------------------
	% Contact formulation
	%--------------------------------------------------------------------	
	\subsection{Contact surface description} \label{s:appndx_surface}
	Considering a 3D body $\cbody$, its boundary $\partial \mathcal{B}$, such as its contact surface $\csurf$, can be described by the mapping
	\begin{equation}
		\bx = \bx(\bxi), \quad \bxi \in \cparam,
	\end{equation}
	which maps a point $\bxi = \left\{ \xi^1, \xi^2 \right\}$ lying in the 2D parameter space $\cparam$ to the surface point $\bx \in \csurf$. Then, a set of covariant tangent vectors on $\csurf$ can be defined as
	\begin{equation} \label{eq:covariant}
		\ba_\alpha := \frac{\partial \bx}{\partial \xi^\alpha}, \quad (\alpha=1,2)
	\end{equation}
	with the normal vector 
	\begin{equation} \label{eq:normal}
		\bn := \frac{\ba_1 \times \ba_2}{||\ba_1 \times \ba_2||}.
	\end{equation}
	With Eqs.~\eqref{eq:covariant} and \eqref{eq:normal}, the contact surface can then be characterized by the surface metric components
	\begin{align}
		a_{\alpha\beta} & = \ba_\alpha \cdot \ba_\beta,  \label{eq:metric}
	\end{align}
	The dual tangent vectors $\ba^\alpha$ defined by the relation $\ba^\alpha \cdot \ba_\beta = \delta^\alpha_\beta$ can be related to the covariant tangent vectors in Eq. \eqref{eq:covariant} by $\ba_\alpha = a_{\alpha\beta} \, \ba^\beta$. To track changes of $\csurf$ during deformation, one chooses a reference configuration $\csurf_0$, where the tangent vectors $\boldsymbol{A}_\alpha$ and the surface metric tensor components $A_{\alpha\beta}$ can be defined as in Equations \eqref{eq:covariant} and \eqref{eq:metric}, respectively.
	%
	%--------------------------------------------------------------------
	% Contact kinematics
	%--------------------------------------------------------------------		
	\subsection{Contact kinematics and contact traction} \label{s:appndx_kinematics}		
	To formulate frictional contact between two bodies $\cbody_1$ and $\cbody_2$, we consider interactions between a so called \textit{slave point} $\bx_k \in \csurf_k$ ($k=$ 1 or 2) and a neighboring \textit{master} surface $\csurf_\ell$ ($l=$ 2 or 1). For the classical \textit{full-pass} contact algorithm (see Laursen and Simo~\cite{laursen1993}) $k$ is either set to 1 or 2. Further, we assume point interaction, i.e., $\bx_k$ can only interact with at most one point $\bx_\ell \in \csurf_\ell$ at a given time. 
	%--------------------------------------------------------------------
	\begin{figure}[H]
		\centering
		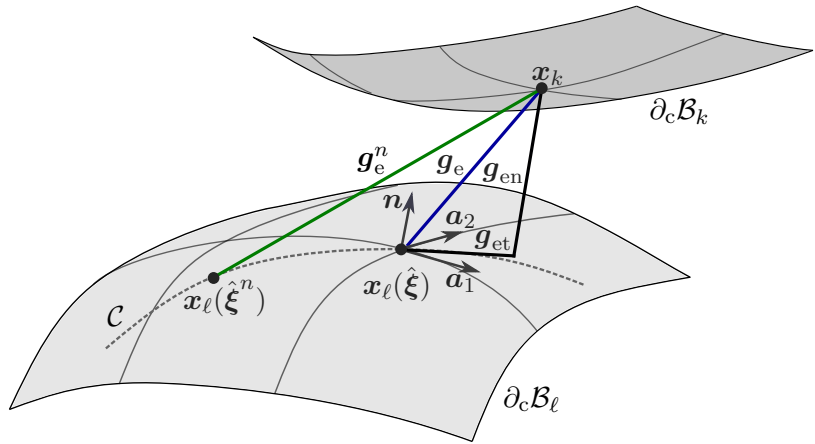
		\caption{Frictional contact kinematics at current time $t_{n+1}$: current slave point $\bx_k$, current position of the previous interacting point $\displaystyle \bx_\ell(\hat{\bxi}^n)$, current interacting point $\bx_\ell(\hat{\bxi})$, the current elastic gap vector $\bge$ and its components, previous elastic gap $\bge^n$, and the sliding path $\mathcal{C}$. Adopted and modified from Ref.~\cite{duong2019}.}
		\label{img:unified_formulation}
	\end{figure}
	%--------------------------------------------------------------------
	In the following, for the sake of conciseness, all variables without superscript $n$ are evaluated at the current time $t_{n+1}$, if not stated otherwise.
	To characterize the interaction between the two bodies we consider a penalty formulation, with the penalty parameter $\epsilon$ (considered equal in normal and tangential direction). In order to determine the contact traction at the current time step, according to the contact formulation of Ref.~\cite{duong2019}, a so-called interacting (elastic) gap vector $\bge(\hat{\bxi})$ is introduced (see \fig{img:unified_formulation}). This gap vector is defined between the current slave point $\bx_k$ and the so-called current interacting point $\bx_\ell(\hat{\bxi})$ on the master surface $\csurf_\ell$ (defined below), i.e.
	\begin{equation}
		\bge(\hat{\bxi}) := \bx_k - \bx_\ell(\hat{\bxi}). \label{eq:elastic_gap}
	\end{equation}
	The current gap vector can be further decomposed into a tangential and normal component $\bge(\hat{\bxi}) = \bgne + \bgte$, with
	\begin{equation}
		\begin{aligned}
			\bgne(\hat{\bxi}) & := \left( \bn \otimes \bn \right) \, \bge,\\
			\bgte(\hat{\bxi}) & := \left( \ba_\alpha \otimes \ba^\alpha \right) \, \bge.
		\end{aligned}
	\end{equation}
	According to the penalty formulation, the total (normal and tangential) frictional contact traction is proportional to the interacting gap vector $\bge(\hat{\bxi})$, according to
	\begin{equation}
		\ctc = \epsilon \, \bge,
	\end{equation}
	which follows from using the contact potential $W:=\frac{1}{2}\epsilon \, \bge \cdot \bge$. At initial contact, the interacting point $\bx_\ell(\hat{\bxi})$ is equal to the closest projection point of $\bx_k$.
	During sticking, the current interacting point is equal to the previous interacting point $\hat{\bxi}^n$.
	Therefore, for sticking, the current contact gap vector $\bge$ is determined from Eq.~\eqref{eq:elastic_gap} with $\hat{\bxi} = \hat{\bxi}^n$.
	During sliding, the current interacting point $\hat{\bxi}$ is the solution of the kinematic constraint equation
	\begin{equation}
		\boldsymbol{f}_{\bg}(\hat{\bxi}) := \bgte - \bgtemax = \mathbf{0},
	\end{equation}
	in the current configuration (see Ref.~\cite{duong2019}). $\bge$ then follows from Eq.~\eqref{eq:elastic_gap}. The critical value during sliding $\bgtemax$ is defined by the chosen friction law. For example, for Coulomb's friction, it is defined as
	\begin{equation}
	\bgtemax = \mu \lv \bgne \rv \btau,
	\end{equation}
	where $\btau$ denotes the sliding direction and can be computed by projecting the previous interacting gap $\bge^n$ onto the tangent plane at the current interaction point $\bx_\ell(\hat{\bxi})$, i.e.
	\begin{equation}
	\btau = \frac{\left( \ba_\alpha \otimes \ba^\alpha \right) \, \bge^n}{\lv \left( \ba_\beta \otimes \ba^\beta \right) \, \bge^n \rv},
	\end{equation}
	where $\ba_\alpha,\,\ba^\alpha,$ are evaluated at $\bx_\ell(\hat{\bxi})$.

	The computation of the friction coefficient $\mu$ in Eq.~\eqref{eq:phi} requires the knowledge of the accumulated sliding distance $\gs$. Here, we approximate Eq.~\eqref{eq:gs} by accumulating the distance from the initial interacting point to the current interacting point, i.e.
	\begin{equation}
		\gs^{n+1} \approx \sum_{i=1}^{n+1} \lv \bx_\ell\left( \hat{\bxi}^i \right) - \bx_\ell\left( \hat{\bxi}^{i-1} \right) \rv.
	\end{equation}
	%
	%--------------------------------------------------------------------
	% FEM
	%--------------------------------------------------------------------	
	\subsection{Finite element discretization} \label{s:appndx_fem}
	For the finite element formulation, the contact bodies are discretized into $n_\mathrm{el}$ elements. The geometry of an element $e$ in the current configuration can be interpolated from the positions of the elemental nodes, or control points $\mathbf{x}_e$, as
	\begin{equation}
		\bx = \bN_e(\bxi) \, \mathbf{x}_e, \quad \bxi \in \cparam,
	\end{equation}
	where $\bN_e := \left[ N_1\bI, N_2\bI, ..., N_{n_\mathrm{el}}\bI \right]$ denotes the element shape function array. Following the 3D contact enrichment approach of Corbett and Sauer~\cite{corbett2014,corbett2015}, the bulk of the bodies is discretized by linear elements with standard Lagrange interpolation following Sauer~\cite{sauer2011b}, while the contact surfaces are discretized by non--uniform rational B-Splines (NURBS) interpolation (see, e.g., Hughes~\etal~\cite{hughes2005}). The NURBS basis functions can be computed element-wise by employing the B\'{e}zier extraction operator $\bC^e$ of Borden~\etal~\cite{borden2011} and the Bernstein polynomials $\bB$. Then, the shape function of control point $A$ can be computed by
	\begin{equation}
		N_A(\xi^1, \xi^2) = \frac{w_A \hat{N}_A^e(\xi^1,\xi^2) }{\sum_{A=1}^{n} w_A \hat{N}_A^e(\xi^1,\xi^2)},
	\end{equation}
	with the corresponding weight $w_A$, and the B-spline basis functions
	\begin{equation}
		\begin{aligned}
			\hat{\bN}^e				 & = \{ \hat{N}_A^e \}^{n_e}_{A=1}, \\
			\hat{\bN}^e(\xi^1,\xi^2) & = \bC^e_{\xi^1} \bB(\xi^1) \otimes \bC^e_{\xi^2} \bB(\xi^2).
		\end{aligned}
	\end{equation}
	The finite element forces acting on the slave surface element $e$ and the master surface element $\hat{e}$ are then given by
	\begin{equation}
	\begin{aligned}
		\bff_c^e		 & := \int_{\Gamma^e} \bN_e^\mathrm{T} \, \ctc \dd a  \quad \text{(slave)},\\
		\bff_c^{\hat{e}} & :=-\int_{\Gamma^{\hat{e}}} \bN_{\hat{e}}^\mathrm{T} (\hat{\bxi}) \, \ctc \dd a \quad \text{(master)},
	\end{aligned}	
	\end{equation}
	where $\Gamma$ denotes the surface element domains on the corresponding surfaces. As NURBS elements are at least $C^1$ continuous over the element boundaries, this gives us an accurate representation of the geometry, as well as an efficient and stable framework for contact computations.	
%
%--------------------------------------------------------------------
% Convergence
%--------------------------------------------------------------------
\section{Derivation of the new analytical model with modified \\ Coulomb's friction} \label{s:appndx_ana_new}
Given the modified Coulomb's law in \sect{s:model_modcoulomb}, a new analytical model can be derived. In general, the torque $T$, as a function of the rotation angle $\theta$, can be computed analytically, as
\begin{equation}
	T(\theta) =  2 \pi \int_{r} r^2 \cdot \sigzt(r,\theta,\gs) \dd r.
\end{equation}
Due to symmetry, the tangential traction component $\sigzt$ is distributed radially symmetric along the BII, while the radial traction component $\sigma_{rz}$ is zero. With the definition of the critical radius $c$ from Eq.~\eqref{eq:rcrit} we have $c(\thetalin) = R$ and
\begin{equation} \label{eq:ana1}
	\sigzt(\thetalin,c(\thetalin)) := \ttmax.
\end{equation}
Assuming bone and implant to be linear elastic bodies, we know that for sticking, the tangential traction will be proportional to the applied rotation angle $\theta$ and the current radius $r$, i.e.
\begin{equation} \label{eq:ana2}
	\sigzt^\text{stick}(\theta,r) = \lambda \theta r, \quad \lambda = \text{const}, 
\end{equation}
where $\lambda$ is a constant stress per length. Thus, the torque for the sticking region becomes
\begin{equation} \label{ana3}
	T(\theta) = 2 \pi \int^{R}_{0}  r^2 \cdot \lambda \theta r \dd r = \frac{1}{2}\pi R^4 \lambda \theta, \text{ for } \theta \le \thetalin.
\end{equation}
If we calculate the slope of the torque at $\theta = \thetalin$ and $c(\thetalin)$, we get
\begin{align} \label{eq:ana4}
	\frac{\dd T(\thetalin, R)}{\dd \thetalin} = \frac{1}{2}\pi R^4 \lambda.
\end{align}
	%--------------------------------------------------------------------
\begin{figure}[H]		
	\centering
	\input{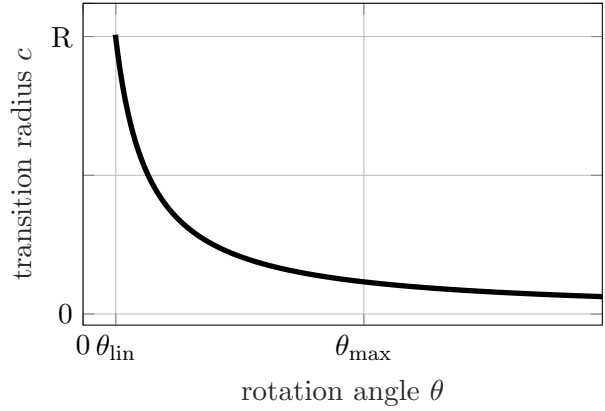}
	\caption{New analytical model: Critical radius for stick/ slip transition $c$ as a function of the imposed rotation angle $\theta$ from Eq.~\eqref{eq:rcrit}.}
	\label{img:rcrit}
\end{figure}
%--------------------------------------------------------------------
We also know that for linear elasticity
\begin{equation} \label{eq:ana5}
	\frac{\dd T(\theta)}{\dd \theta} = C, \quad \theta\le\thetalin,
\end{equation}
where $C$ is the effective shear stiffness of the system, that is approximately proportional to the shear modulus of bone. If we equate Eqs.~\eqref{eq:ana4} and~\eqref{eq:ana5}, we obtain
\begin{equation} \label{eq:ana6}
	\lambda = \frac{2C}{\pi R^4},
\end{equation} 
and inserting $\lambda$ into Eq.~\eqref{eq:ana2} yields
\begin{equation} \label{eq:ana7}
	\sigzt^\text{stick}(\theta,r) = \frac{2C}{\pi R^4} \theta r . 
\end{equation}
Applying the definition in Eq.~\eqref{eq:ana1} to Eq.~\eqref{eq:ana7} yields
\begin{equation}
	\sigzt^\text{stick}(\thetalin,R) = \frac{2C}{\pi R^4} \thetalin R  = \ttmax,
\end{equation}
i.e.
\begin{equation}
	\frac{2C}{\pi R^3}=\frac{\ttmax}{\thetalin}.
\end{equation}
Together with Eq.~\eqref{eq:rcrit} we finally obtain the tangential traction component for sticking
\begin{equation} \label{eq:ana8}
	\sigzt^\text{stick}(\theta,r) = \theta r \frac{\ttmax}{\thetalin R}=\frac{r}{c(\theta)} \ttmax.
\end{equation} 
In the sliding region $r \ge c$, the tangential traction follows the modified Coulomb's law 
\begin{equation} \label{eq:ana9}
	\sigzt^\text{slide}=\mu p,
\end{equation}
with $\mu = \mu(\gs)$ from Eq.~\eqref{eq:mod_mu}. Here, we assume $p$ to be homogeneously distributed along the contact surface. As long as a part of the interface remains sticking 
\begin{equation}
	\gs = 0, \quad \text{otherwise} \quad \gs = r \left( \theta - \thetamax + \thetalin \right).
\end{equation} 
The tangential contact traction as a function of the rotation angle can then be summarized by combining Eq.~\eqref{eq:ana8} and Eq.~\eqref{eq:ana9} to Eq.~\eqref{eq:new_ana}.
%
%--------------------------------------------------------------------
% Convergence
%--------------------------------------------------------------------
\section{Mesh and load step sensitivity} \label{s:appndx_conv}
To analyze the convergence behavior, three different finite element meshes were constructed, denoted \textit{coarse}, \textit{medium}, and \textit{fine} with 3,390, 12,354, and 47,130 degrees of freedom, respectively. In addition, different load step sizes [\ang{0.1}, \ang{0.05}, \ang{0.02}, \ang{0.01}, \ang{0.005}, \ang{0.004}] were investigated, corresponding to a number of load steps of [100, 200, 500, 1,000, 2,000, 2,500], respectively. For the parameters, data set 1 with $\muk=0.4$ was chosen. To compare the different setups, we define the mean relative torque error
\begin{equation} 
	\errrel = \underset{\theta\in\left[0,\ang{10}\right]}{\mathrm{mean}} \left( \lv \frac{\Texp(\theta)-T(\theta)}{\Tmax} \rv \right), \label{eq:erel}
\end{equation}
where here $T_\mmax$ is the maximum torque obtained by the numerical solution.	

\fig{img:mre_exp} shows the convergence behavior of the different meshes. It can be seen that $\errmp$ reaches its limit for all meshes after 1,000 load steps to 0.0175, 0.0171, and 0.017, respectively. This is also the case for the mean percentage error shown in \tab{tab:conv}, with its lowest value of \SI{2.176}{\percent} for the fine mesh and the highest value of \SI{2.241}{\percent} for the coarse mesh. In addition, the error is increasing with the number of load steps for the coarse mesh. This stems from the coarse resolution of the peak for larger load steps and thus leading to the torque values to be closer to the experimental data.
It should be noted, that the mesh size has a small effect on the outcome of the parameter study and thus, both $\errmp$ and $\errrel$ can be further minimized by performing a separate parameter estimation for each mesh.\\
%--------------------------------------------------------------------
\begin{minipage}[b]{0.47\linewidth}
	\vspace{12mm}
	\begin{table}[H]	
		\begin{tabular}{|c||c|c|c|}
			\hline 
			load step & coarse &  medium & fine \\ 
			\hline \hline
			\ang{0.10} & 0.02233 & 0.02195 & 0.02186 \\
			\ang{0.05} & 0.02242 & 0.02197 & 0.02185 \\
			\ang{0.02} & 0.02239 & 0.02191 & 0.02177 \\
			\ang{0.01} & 0.02240 & 0.02190 & 0.02176 \\
			\ang{0.005}& 0.02241 & 0.02190 & 0.02176 \\
			\ang{0.004}& 0.02241 & 0.02190 & 0.02176 \\
			\hline
		\end{tabular}
		\vspace{17mm}
		\caption{Mesh sensitivity: Mean percentage error $\errmp$ according to Eq.~\eqref{eq:mpe} for different configurations of data set 1 ($\muk=0.4$).}
		\label{tab:conv}
	\end{table}
\end{minipage}
\hfill
\begin{minipage}[b]{0.51\linewidth}
	\begin{figure}[H]
\definecolor{mycolor1}{rgb}{0.00000,1.00000,1.00000}%
\begin{tikzpicture}

\begin{axis}[%
width=2.25in,
height=1.7745in,
at={(0.758in,0.481in)},
scale only axis,
scaled y ticks= false,
xmin=0,
xmax=2500,
xlabel style={font=\color{white!15!black}},
xlabel={load steps},
ymin=0.0169,
ymax=0.0178,
xmajorgrids,
ymajorgrids,
ylabel style={font=\color{white!15!black}},
yticklabel style={/pgf/number format/fixed, /pgf/number format/precision=4, /pgf/number format/fixed zerofill},
ylabel={$e_T^\mathrm{rel}$},
axis background/.style={fill=white},
axis x line*=bottom,
axis y line*=left,
legend style={legend cell align=left, align=left, draw=white!15!black}
]
\addplot [color=lightgray, line width=2.0pt]
  table[row sep=crcr]{%
100	0.01769\\
200	0.01757\\
500	0.01747\\
1000	0.01746\\
2000	0.01746\\
2500	0.01746\\
};
\addlegendentry{coarse}

\addplot [color=gray, line width=2.0pt]
  table[row sep=crcr]{%
100	0.01742\\
200	0.01724\\
500	0.01711\\
1000	0.01709\\
2000	0.01708\\
2500	0.01708\\
};
\addlegendentry{medium}

\addplot [color=black, line width=2.0pt]
  table[row sep=crcr]{%
100	0.01736\\
200	0.01716\\
500	0.01702\\
1000	0.01699\\
2000	0.01697\\
2500	0.01697\\
};
\addlegendentry{fine}

\end{axis}

%\begin{axis}[%
%width=5.833in,
%height=4.375in,
%at={(0in,0in)},
%scale only axis,
%xmin=0,
%xmax=1,
%ymin=0,
%ymax=1,
%axis line style={draw=none},
%ticks=none,
%axis x line*=bottom,
%axis y line*=left,
%legend style={legend cell align=left, align=left, draw=white!15!black}
%]
%\end{axis}
\end{tikzpicture}%
		\caption{Mesh sensitivity: Mean relative error $\errrel$ according to Eq.~\eqref{eq:erel} for different configurations of data set 1 ($\muk=0.4$).}
		\label{img:mre_exp}
	\end{figure}
\end{minipage}
%
%--------------------------------------------------------------------
%  BIBLIOGRAPHY
%--------------------------------------------------------------------
\bibliographystyle{habbrv}
\bibliography{./Bibliography_all}

\end{document}